\documentclass[aps, prb, twocolumn,amsmath,amssymb,floatfix, reprint, longbibliography]{revtex4-2}
\usepackage{graphicx}
\usepackage{svg}
\usepackage{times}
\usepackage{dcolumn}
\usepackage{hyperref}
\usepackage{chemformula}
\hypersetup{colorlinks, allcolors=blue}
\usepackage{lipsum}
\usepackage{comment}

\begin{document}
\title{Multiband strong-coupling superconductors with spontaneously broken time-reversal symmetry}
\author{Niels Henrik Aase}
\affiliation{\mbox{Center for Quantum Spintronics, Department of Physics, Norwegian University of Science and Technology, NO-7491 Trondheim, Norway}}
\author{Kristian M{\ae}land}
\affiliation{\mbox{Center for Quantum Spintronics, Department of Physics, Norwegian University of Science and Technology, NO-7491 Trondheim, Norway}}
\author{Asle Sudb{\o}}
\email[Corresponding author: ]{asle.sudbo@ntnu.no}
\affiliation{\mbox{Center for Quantum Spintronics, Department of Physics, Norwegian University of Science and Technology, NO-7491 Trondheim, Norway}}

\begin{abstract}
We study superconducting three-band systems within strong-coupling Eliashberg theory. In particular, we search for phase-frustrated superconducting systems with spontaneous time-reversal symmetry breaking (TRSB) states.
The emergence of TRSB states in multiband systems has so far been studied using microscopic weak-coupling BCS theory or more phenomenological effective field theories such as multi-component Ginzburg-Landau theories.
For systems with three disjoint Fermi surfaces whose electrons experience interactions mediated by phonons, we present a microscopic analysis showing that TRSB states also exist within a strong-coupling microscopic theory.
The systems we consider have sizable electron-phonon couplings, putting them into the strong-coupling regime. They are thus a fitting description for strong-coupling materials such as some of the iron pnictides.
Moreover, as the TRSB states are challenging to find numerically, we calculate the free energy of multiband systems within strong-coupling theory and make explicit use of it to pin down the TRSB states’ elusive nature.
Since Eliashberg theory is well incorporated with first-principles calculations, our strong-coupling approach might help facilitate a more efficient search for candidate materials that can exhibit TRSB. 
\end{abstract}

\maketitle
\section{Introduction}
Superconductors represent a remarkable quantum state of matter where photons acquire a mass through the Higgs mechanism (the Meissner effect) when metals are cooled below their critical temperature.
When combined with spontaneous time-reversal symmetry breaking (TRSB), this extraordinary quantum state exhibits even more fascinating features due to the interplay between multiple broken symmetries.
TRSB may occur in multiband superconductors where several bands (more than two) cross the Fermi surface.
Many multiband superconductors exist \cite{Nagamatsu2001, Kamihara2006, Kamihara2008, Dvir2018, Fernandes2022}, including several relatively high-temperature superconductors \cite{Rotter2008, Ren2008, Shibauchi2009, Nakayama2009}, indicative of some strong pairing glue between electrons being operative.
So far, theoretical studies of TRSB in multiband superconductors have been carried out using either Ginzburg-Landau theory \cite{Ng2009, Carlstrom2011, Bojesen2013, Garaud2013, Bojesen2014, Lin2016, Garaud2017} or a more microscopic weak-coupling BCS theory \cite{Stanev2010, Tanaka2010, Wilson2013}.
Given that many of the TRSB systems of interest are suspected to be strong-coupling superconductors, this motivates similar studies within the framework of a strong-coupling approach to superconductivity.  

An elegant approach to strong-coupling Eliashberg theory was recently realized \cite{Protter2021}. It utilizes the imaginary-time functional integral formalism \cite{Negele1988, Atland2010}, typically employed to derive effective field theories of interacting condensed matter systems. The original microscopic theory for superconductivity \cite{Bardeen1957} is easily derived using these field-theory techniques. Such a treatment sets the stage for systematic studies of higher-order corrections within the BCS paradigm and has been extensively used. However, the retardation effects required to obtain the physical picture of electrons with attractive interactions mediated by lattice vibrations necessitate a dynamical interaction which is absent in BCS theory. This is remedied in Eliashberg theory \cite{Eliashberg1960, Eliashberg1961, Marsiglio2020}, with the additional benefit that the strong-coupling regimes \cite{Scalapino1966} can be probed. 

While Migdal's theorem \cite{Migdal1958} may pose some limitations on strong-coupling Eliashberg theory, recent studies have found ways of accounting for it by means of vertex corrections \cite{Schrodi2020, Schrodi2021}. Moreover, there is also a case to be made that surpassing the Migdal limit is not necessarily a hindrance to capturing the relevant physics at play \cite{Bauer2011, Chubukov2020}.
Regardless of the precise nature of the breakdown of Eliashberg theory, the practice of pushing it far into the strong-coupling regime is widespread, and the results obtained often corroborate experimental findings in materials with a sizable electron-phonon coupling. So with the advantages of Eliashberg theory and especially its applicability to strong-coupling systems, Ref.\ \cite{Protter2021} represents a significant step forward as it derives Eliashberg theory using a functional-integral approach, thus opening the door to more sophisticated field-theoretic treatments previously reserved mainly for BCS theory. 

The novel functional-integral approach to Eliashberg theory has already been employed in several models, e.g.\ graphene multilayers \cite{Chou2022}, quantum-critical models \cite{Zhang2022, Yuzbashyan2022b}, systems that are superconducting despite experiencing instantaneous repulsion \cite{Dalal2023} and systems of coexisting antiferromagnetism and superconductivity \cite{Lundemo2023}. In this approach, the Eliashberg equations for the various systems are derived as stationary point conditions in the action.
At a stationary point, the action can be used to determine the free energy of the system, so in systems with more than one solution to the Eliashberg equations, the free-energy perspective provides a simple selection criterion for such cases, namely that the stationary point with the lowest free energy should be chosen.
This selection criterion is particularly useful in systems where several states may solve the Eliashberg equations. Such an analysis has already been put forth to understand the breakdown of Eliashberg theory \cite{Yuzbashyan2022}, but other avenues remain unexplored.

An uncharted area of interest in this regard is the field of multiband superconductors. Since the discovery of superconductivity in \ch{MgB_2} \cite{Nagamatsu2001} with a pronounced multigap character \cite{Szabo2001, Tsuda2001}, as well as the superconducting iron pnictides \cite{Kamihara2006, Kamihara2008}, multiband superconductors have been intensely researched theoretically \cite{Kuroki2008, Mazin2008, Silaev2011, Hirschfeld2011, Wilson2013, Yuan2014, Galteland2016, Maiani2022, Fernandes2022}. In many cases, the theoretical treatments are based on the extension of BCS theory to multiband systems \cite{Suhl1959, Kondo1963}.
Applying these to a two-band model for ${\mathrm{LaFeAsO}}_{1\ensuremath{-}x}{\mathrm{F}}_{x}$ \cite{Mazin2008}, it was found that repulsive, rather than attractive, interactions may drive the system into a superconducting $s^\pm$-wave phase, where both gaps have $s$-wave symmetry and are of opposite sign.

Despite its simplicity, the $s^\pm$-phase explains quite well the anomalous experimental findings in multiband systems \cite{Bang2017}.
However, in three-band systems, a novel superconducting phase may emerge, where the phase differences of the various order-parameter components are not necessarily multiples of $\pi$ \cite{Ng2009, Stanev2010, Tanaka2010}.
These states break time-reversal symmetry (TRS) and have been shown to exhibit a number of exotic phenomena, including unconventional vortex physics \cite{Garaud2013, Gillis2014} and anomalous metallic states above $T_{\mathrm{c}}$ \cite{Bojesen2014, Grinenko2021}. 
The latter has been suggested as a possible explanation for the magnetic memory observed in \ch{TaS_2} above $T_\mathrm{c}$ \cite{Persky2022}.
One of the consequences of breaking TRS is the emergence of local spontaneous magnetic fields \cite{Garaud2014, Lin2016}, which has been experimentally measured using the muon spin relaxation technique \cite{Grinenko2020}.

As previously mentioned, in most studies, multiband states with broken TRS are treated using BCS theory or by effective field theories (e.g., Ginzburg-Landau theories) derived from BCS theory following the original calculation of Gor'kov \cite{Gorkov1959, Zhitomirsky2004}. Monte Carlo simulations have been used to perform computations using such effective theories beyond the mean-field level \cite{Bojesen2013, Bojesen2015}.
However, as mentioned earlier, these multiband systems have so far not been studied in the strong-coupling regime even though the archetypical systems one might expect TRSB in, namely the superconducting iron pnictides, often involve strong coupling \cite{Stewart2011}. 
It is thus of interest to examine the fate of TRSB states in strong coupling, and we will do so using a strong-coupling functional-integral approach, which has a saddle point equivalent to the Eliashberg theory of superconductivity. 
Moreover, the ease with which first-principles calculations can be incorporated into such a strong-coupling theory can help facilitate the search for candidate materials that may host TRSB states.  

The structure of the paper is as follows. In Sec.\ \ref{sec:funcint}, by using the functional-integral formalism, we derive the multiband Eliashberg equations and the free energy of superconducting multiband systems. In light of this, we discuss the prerequisites for having spontaneous TRSB states in strong coupling in Sec.\ \ref{sec:GPA}. While these two sections are general and may apply to several models, we then consider a microscopic model in \ref{sec:micro} and derive the Eliashberg equations specific to this model.
In Sec.\ \ref{sec:results}, we first demonstrate the spontaneous breaking of TRS in strong coupling and subsequently elucidate its origin. We discuss the experimental consequences of our results and some of the advantages of employing a strong-coupling theory. The conclusions are finally presented in Sec.\ \ref{sec:conclusion}.

\section{The functional-integral formalism and multiband Eliashberg theory}\label{sec:funcint}
The purpose of this section is to derive the free energy for multiband systems with dynamic interactions as well as the associated Eliashberg equations. The latter has been carried out in the single-band case Ref.\ \cite{Protter2021,Lundemo2023}, and we will generalize these results to multiband systems. Since the derivation (and the notation) will be similar to that of Ref.\ \cite{Protter2021}, we mainly focus on the parts that are different in this paper and refer to the original work for some of the more technical details. Furthermore, we stress that multiband systems have been studied using Eliashberg theory \cite{Shulga1998, Nicol2005, Ummarino2009, Ummarino2013}, so while our derivation of the Eliashberg equations using the functional integral-formalism is novel, the end result is not. With this approach, we also derive the free energy, which will subsequently be used to discover TRSB states.

Within the imaginary time functional-integral formalism, the dynamics of a multiband system of fermions is governed by the action $S$. If the interactions in the system are density-density interactions, one can perform Hubbard-Stratonovich (HS) decoupling in three different channels - Cooper, density, and exchange \cite{Protter2021}. In the interest of studying superconductivity and renormalization effects, we will decompose the interaction into the Cooper and density channels and neglect the exchange channel. Thus, when we write down the action, we write the interacting part in a suggestive form, preparing for the HS decouplings to come. Assuming that the interactions only depend on relative coordinates (and thus only on relative momentum), the action is
\begin{align}
        &S[\bar{\psi}, \psi] =  \sum_k \sum_{\sigma, i} \bar{\psi}_{\sigma i}(k) [-i\omega_n +\xi_i(\mathbf{k})]\psi_{\sigma i}(k) \nonumber \\
        &+ \frac{1}{2}\sum_{k, k'} \sum_{\sigma, i, j} \bar{\psi}_{\sigma i}(k)\psi_{\sigma i}(k) V_{ij}(k-k')\bar{\psi}_{\sigma j}(k')\psi_{\sigma j}(k') \nonumber \\
        &- \sum_{k, k'}\sum_{i, j} \bar{\psi}_{\uparrow i}(k)\bar{\psi}_{\downarrow i}(-k) V_{ij}(k-k')\psi_{\downarrow j}(-k')\psi_{\uparrow j}(k'),
    \label{starting_point_Eliash}
\end{align}
where we introduced the shorthand notation $k = (\omega_n, \mathbf{k})$ with $\omega_n$ being the fermionic Matsubara frequency $\omega_n = \pi(2n+1)/\beta$ and $\beta = 1/T$ being inverse temperature, where we use natural units $\hbar=c=k_{\mathrm{B}} = 1$. 
$\psi_{\sigma i}(k)$ ($\bar{\psi}_{\sigma i}(k)$) is the Grassmann field representing the annihilation (creation) operator of an electron with spin $\sigma$ belonging to band $i$. $\xi_i(\mathbf{k})$ is the dispersion relation $\varepsilon_i(\boldsymbol{k})$ of band $i$, relative to the chemical potential $\xi_i(\mathbf{k}) = \varepsilon_i(\mathbf{k})-\mu$. We have also assumed the Cooper channel to only have pairings with opposite momentum. 

For the interaction $V_{ij}(k-k')$ to be separable in the band indices as in Eq.\ \eqref{starting_point_Eliash}, we implicitly assume that the Fermi surfaces are disjoint. Multiple disjoint Fermi surfaces may originate with having several bands crossing the Fermi energy or having a band crossing it several times, creating disjoint Fermi pockets.
Several superconducting systems of current interest fit this description, for example, the superconducting iron pnictides and transition metal (di)chalcogenides \cite{Chubukov2009, Stewart2011, DeLaBarrera2018, Wickramaratne2020, Fernandes2022}.
In all these cases, hybridization effects stemming from the self-energy are absent, supposing that the Fermi surfaces are sufficiently well separated. 

With the path integral measure $\mathcal{D} \psi\mathcal{D} \bar{\psi}$, the grand canonical partition function $\mathcal{Z}$ at fixed chemical potential is given by $\mathcal{Z} = \int \mathcal{D} \psi\mathcal{D} \bar{\psi}\mathrm{e}^{-S[\bar{\psi}, \psi]} = \mathrm{e}^{-\beta F}$ , where $F$ is the free energy (or more precisely, $F$ is the grand potential, which we take to be equal to the free energy when fixing the chemical potential) of the system. So, in keeping with conventional HS decoupling, we may write
\begin{align}
    1 &= \int \mathcal{D}\bar{\Phi}\mathcal{D}\Phi\mathcal{D}\Sigma^\uparrow\mathcal{D}\Sigma^\downarrow \nonumber \\
    &\mathrm{e}^{-\sum_{k, k'}\sum_{i,j}V^{-1}_{ij}(k-k') [\bar{\Phi}_i(k) \Phi_j(k') + \frac{1}{2}\Sigma_i^\sigma(k)\Sigma^\sigma_j(k')]},
    \label{one_HS}
\end{align}
where a summation over repeated spin indices is implied. The inverse interaction $V_{il}^{-1}(k_1 - k)$ is a function which obeys the relation
\begin{equation}
    \sum_k \sum_l V_{il}^{-1}(k_1 - k) V_{lj}(k-k_2) = \delta_{k_1k_2}\delta_{ij}. 
    \label{inverse_relation}
\end{equation}
We discuss the inverse interaction in detail later.
$\bar{\Phi}$, $\Phi$ and $\Sigma^\sigma$ are shifted to get rid of the quartic terms in Eq.\ \eqref{starting_point_Eliash}, however the shifts are slightly more complicated than those of Ref.\ \cite{Protter2021}
\begin{subequations}
\begin{align}
    \Phi_i(k) &\rightarrow \Phi_i(k) - \sum_{k', j} V_{ij}(k-k')\psi_{j\downarrow}(-k')\psi_{j\uparrow}(k')\\
        \bar{\Phi}_i(k) &\rightarrow \bar{\Phi}_i(k) - \sum_{k', j} V_{ji}(k'-k)\bar{\psi}_{j\uparrow}(k')\bar{\psi}_{j\downarrow}(-k') \\
        \Sigma^\sigma_i(k) &\rightarrow \Sigma^\sigma_i(k) + i\sum_{k', j} V_{ij}(k-k')\bar{\psi}_{j\sigma}(k')\psi_{j\sigma}(k').
\end{align}
\label{shifted_fields}
\end{subequations}
The imaginary shift of $\Sigma^\sigma_i(k)$ is due to the fact that in the density channel, the interaction is repulsive. Inserting Eq.\ \eqref{one_HS} into Eq.\ \eqref{starting_point_Eliash}, and subsequently employing the linear shifts in Eq.\ \eqref{shifted_fields} while making repeated use of Eq.\ \eqref{inverse_relation}, straightforwardly yields
\begin{widetext}
\begin{align}
        &S[\bar{\psi}, \psi, \bar{\Phi}, \Phi, \Sigma] = 
        \sum_k \sum_{\sigma, i} \bar{\psi}_{\sigma i}(k) [-i\omega_n +\varepsilon_i(\mathbf{k}) + i\Sigma^\sigma_i(k)]\psi_{\sigma i}(k) 
        - \sum_{k, i} [\bar{\psi}_{\uparrow i}(k) \bar{\psi}_{\downarrow i}(-k) \Phi_i(k) + \bar{\Phi}_i(k)\psi_{\downarrow i}(-k)\psi_{\uparrow i}(k) ] \nonumber \\
        &+ \sum_{k, k'}\sum_{i,j}V^{-1}_{ij}(k-k') [\bar{\Phi}_i(k) \Phi_j(k') + \frac{1}{2}\Sigma_i^\sigma(k)\Sigma^\sigma_j(k')].
    \label{Eliash_post_HS}
\end{align}
The action in Eq.\ \eqref{Eliash_post_HS} is diagonal in the electron bands and bilinear in the fermionic fields, such that we can employ the Nambu spinors $\Psi_i(k) = (\psi_{\uparrow i}(k), \bar{\psi}_{\downarrow i}(-k))^{\mathrm{T}}$ and $\bar{\Psi}_i(k) = (\bar{\psi}_{\uparrow i}(k), \psi_{\downarrow i}(-k))$ to express $S$ as
\begin{align}
        &S[\bar{\Psi}, \Psi, \bar{\Phi}, \Phi, \Sigma] = \sum_k \sum_{i} \bar{\Psi}_{i}(k) [-\mathcal{G}_i^{-1}(k)]\Psi_{i}(k)
        + \sum_{k, k'}\sum_{i,j}V^{-1}_{ij}(k-k') [\bar{\Phi}_i(k) \Phi_j(k') + \frac{1}{2}\Sigma_i^\sigma(k)\Sigma^\sigma_j(k')],
    \label{Eliash_post_HS_nambu}
\end{align}
\end{widetext}
where the inverse Green's function $\mathcal{G}_i^{-1}$ is band dependent and given by
\begin{equation}
    \mathcal{G}_i^{-1} = \begin{pmatrix}
        G^{-1}_{i, \uparrow}(k) & \Phi_i(k) \\
        \bar{\Phi}_i(k) & - G^{-1}_{i, \downarrow}(-k)
    \end{pmatrix}
    \label{G_matrix_def}
\end{equation}
with $G^{-1}_{i, \sigma}(k) = G_{0,i}^{-1}(k) - i\Sigma_i^\sigma(k)$ and $G_{0,i}^{-1}(k) = i\omega_n - \xi_i(\mathbf{k})$. The first term in Eq.\ \eqref{Eliash_post_HS_nambu} is block-diagonal in the bands and can thus be treated in the same manner as in Ref.\ \cite{Protter2021}. The two HS terms contain the coupling between the bands and are thus responsible for the emergent multiband physics. 

To proceed, we integrate out the fermionic sector such that the effective action becomes
\begin{align}
        &S[\bar{\Phi}, \Phi, \Sigma] = -\mathrm{Tr}\ln(-\beta \mathcal{G}^{-1})  \nonumber \\
        &+ \sum_{k, k'}\sum_{i,j}V^{-1}_{ij}(k-k') [\bar{\Phi}_i(k) \Phi_j(k') + \frac{1}{2}\Sigma_i^\sigma(k)\Sigma^\sigma_j(k')],
    \label{Eliash_post_fermion_integral}
\end{align}
where the trace is over momentum, frequency and band indices. To find the mean-field values of the HS fields, we impose stationary point conditions on the action and make use of the identity $\frac{\delta}{\delta\chi}\mathrm{Tr}\ln(f(\chi) = \mathrm{Tr}[f^{-1}(\chi) \frac{\delta}{\delta\chi} f(\chi)]$ to obtain (see Ref.\ \cite{Protter2021} for more details) 
\begin{align}
    0 &=\frac{\delta S[\bar{\Phi}, \Phi, \Sigma]}{\delta \bar{\Phi}_i} = \sum_{k', j} V^{-1}_{ij}(k-k')\Phi_j(k') \nonumber \\
    &- \frac{\Phi_i(k)}{G^{-1}_{\uparrow,i}(k)G^{-1}_{\downarrow,i}(-k) + \bar{\Phi}_i(k)\Phi_i(k)} \label{inverted_eliash_eq_sigma},
\end{align}
which can be inverted by summing over $\sum_{k}\sum_i V_{li}(k'-k)$ on both sides and using Eq.\ \eqref{inverse_relation}. After a suitable relabeling of variables, the gap equation is given as
\begin{equation}
    \Phi_i(k) = \sum_{k',j} V_{ij}(k-k') \frac{\Phi_j(k')}{G^{-1}_{\uparrow,j}(k')G^{-1}_{\downarrow,j}(-k') + \bar{\Phi}_j(k')\Phi_j(k')}.
    \label{eliash_eq_Phi}
\end{equation}
Since the system described by Eq.\ \eqref{starting_point_Eliash} has spin-rotational symmetry, $\Sigma^\uparrow_i(k) = \Sigma^\downarrow_i(k) = \Sigma_i(k)$. The spin index in $G^{-1}_{i, \sigma}(k) = G^{-1}_{i}(k)$ can then be omitted. This simplifies the stationary point conditions $0 =\frac{\delta S[\bar{\Phi}, \Phi, \Sigma]}{\delta \Sigma^\sigma_i}$, which are found analogously to those of $\Phi_i(k)$
\begin{equation}
    i\Sigma_i(k) = \sum_{k',j} V_{ij}(k-k') \frac{G_j(-k')}{G^{-1}_{j}(k')G^{-1}_{j}(-k') + \bar{\Phi}_j(k')\Phi_j(k')}.
    \label{eliash_eq_sigma}
\end{equation} 

Equations.\ \eqref{eliash_eq_Phi} and \eqref{eliash_eq_sigma} are the Eliashberg equations for the HS fields $\Phi$ and $\Sigma$. They can be rewritten to elucidate their connection with the original Eliashberg fields, namely the inverse quasiparticle residue $Z$, the shift of the energy spectrum $\chi$, and the superconducting gap appearing in the electron spectrum $\Delta$.
To do so, we use that $\bar{\Phi}_i(k) = \Phi^*_i(k)$ such that the denominator in Eqs.\ \eqref{eliash_eq_Phi} and \eqref{eliash_eq_sigma} is even. Hence we can obtain distinct equations for the odd and even part of $\Sigma_i(k)$.
Introducing the aforementioned Eliashberg fields by
\begin{align}
    i\Sigma_i(k) &= -i\omega_n + i\omega_nZ_i(k) + \chi_i(k) \label{sigma_trans} \\
    \Delta_i(k) &= \frac{\Phi_i(k)}{Z_i(k)} \label{phi_trans},
\end{align}
it follows that
\begin{align}
    \Theta_i(k) \equiv& G^{-1}_{\uparrow,i}(k)G^{-1}_{\downarrow,i}(-k) + |\Phi_i(k)|^2  \nonumber \\
    =&(\omega_nZ_i(k))^2 + (\xi_i(\mathbf{k}) + \chi_i(k))^2 + |Z_i(k)\Delta_i(k)|^2.
    \label{determinant}
\end{align}
Inserting Eq.\ \eqref{sigma_trans} into Eq.\ \eqref{eliash_eq_sigma} and taking the even (odd) part, we get the Eliashberg equation for $\chi_i(k)$ ($Z_i(k)$)
\begin{align}
    \chi_i(k) &= -\sum_{k',j} V_{ij}(k-k') \frac{\xi_j(\mathbf{k'}) + \chi_j(k')}{\Theta_j(k')} \label{eliash_eq_for_chi} \\
    Z_i(k) &= 1 + \frac{1}{\omega_n} \sum_{k',j}V_{ij}(k-k')\frac{\omega_{n'}Z_j(k')}{\Theta_j(k')}\label{eliash_eq_for_Z}, 
\end{align}
whereas the gap equation becomes
\begin{equation}
    \Delta_i(k) = \frac{1}{Z_i(k)}\sum_{k',j} V_{ij}(k-k') \frac{Z_j(k)\Delta_j(k')}{\Theta_j(k')} \label{eliash_eq_for_Delta}.
\end{equation}
In the case of only one band, Eqs.\ \eqref{eliash_eq_for_chi}, \eqref{eliash_eq_for_Z}, and \eqref{eliash_eq_for_Delta} reduce to the conventional Eliashberg equations \cite{Marsiglio2020}. We note that we will continue to make use of the HS fields ($\Phi$ and $\Sigma^\sigma$) in the analytical calculations as needed, but when stating results we will use the Eliashberg fields as they relate to quantities that are experimentally measurable. 

After integrating out the fermions, the free energy of the system is most easily expressed using the HS fields
\begin{equation}
    \mathrm{e}^{-\beta F} = \int \mathcal{D}\bar{\Phi}\mathcal{D}\Phi\mathcal{D}\Sigma \mathrm{e}^{-S[\bar{\Phi}, \Phi, \Sigma]}.
\end{equation}
By expanding $S$ around its saddle point and neglecting higher-order contributions, one obtains 
\begin{align}
    &\beta F = -\sum_{k}\sum_i \mathrm{ln}(\beta^2 \Theta_i(k))  \nonumber\\
    &+ \sum_{k, k'}\sum_{i,j}V^{-1}_{ij}(k-k') [\bar{\Phi}_i(k) \Phi_j(k') + \frac{1}{2}\Sigma_i^\sigma(k)\Sigma^\sigma_j(k')] \label{free_energy_with_momentum},
\end{align}
where we used that $\mathrm{Tr}\; \mathrm{ln}(\mathcal{G}^{-1}) = \mathrm{ln}\; \mathrm{det} (\mathcal{G}^{-1} )$.

To make use of the free energy in Eq.\ \eqref{free_energy_with_momentum} we need to calculate the inverse interaction $V_{ij}^{-1}(k-k')$. This is not required when deriving the mean-field equations, since one can eliminate $V_{ij}^{-1}(k-k')$, as was done in going from Eq.\ \eqref{inverted_eliash_eq_sigma} to Eq.\ \eqref{eliash_eq_Phi}.
The inverse interaction has only been calculated and used in the free energy for a limited class of systems.
One such interaction is the multiband BCS interaction, where the inverse interaction enters as the matrix inverse of the matrix containing the couplings between different bands \cite{Benfenati2023}.
Only recently, starting with Ref.\ \cite{Protter2021} and subsequently used in, e.g., Refs.\ \cite{Yuzbashyan2022, Zhang2022}, has a feasible way of calculating $V^{-1}$ for $k$-dependent potentials emerged. 
The method is conceptually simple and it is instructive to consider a one-band system. Denoting the operation of taking both the spatial and temporal Fourier transform as $\mathcal{F}$, the inverse interaction can be expressed as
\begin{equation}
    V^{-1}(k) = \mathcal{F}\bigg(\frac{1}{\mathcal{F}^{-1}(V(k))}\bigg).
    \label{loose_def_of_V_inv}
\end{equation}
In writing down Eq.\ \eqref{loose_def_of_V_inv}, we have glossed over some finer details relating to the differences in the temporal and spatial Fourier transforms, but we show in detail in Sec.\ \ref{sec:micro} that this definition of $V^{-1}(k)$ satisfies Eq.\ \eqref{inverse_relation}.
For systems with more than one band, one must also take the matrix inverse, as we will see in Sec.\ \ref{sec:micro}.

\section{Superconducting gaps with global phases and spontaneously broken time-reversal symmetry}\label{sec:GPA}
The complex phases of $\Phi_i(k)$ in the free energy in Eq.\ \eqref{free_energy_with_momentum} are only present in the second term. Determining these phases can be carried out directly by solving the gap equation in Eq.\ \eqref{eliash_eq_Phi}, but since this generally requires numerics, some insight is lost. In one-band systems, however, we may do some simple analytical considerations. For example, it follows immediately that if $\Phi(k)$ has only a global phase $\Phi(k) = |\Phi(k)|\mathrm{e}^{i\theta} \; \forall \; k$, the phases cancel, and one is left with solving the gap equation for the amplitudes $|\Phi(k)|$, where the existence of a nontrivial solution is decided by $V(k-k')$. 
This global $U(1)$ symmetry is spontaneously broken at the onset of superconductivity.
One might also consider states with a $k$-dependent phase that solve the complex gap equations. However, as proved in Ref.\ \cite{Yuzbashyan2022a} by using a spin chain representation, for systems with phonon-mediated interactions, the global minimum of the free energy does indeed only have a global phase. This property for single-band systems motivates the global phase ansatz (GPA) for multiband systems, namely that at the global energy minimum, the gaps belonging to each band only have a global phase
\begin{equation}
    \Phi_i(k) = |\Phi_i(k)|\mathrm{e}^{i\theta_i} \; \forall \; k
    \label{GPA}
\end{equation}
Note that the global phase of each gap need not be the same, $\theta_i \neq \theta_j$. We will often employ the shorthand notation $\theta_{ij} \equiv \theta_i - \theta_j$.

The GPA is further supported by the fact that in systems with phonon-mediated interactions we have considered, we have not observed any numerical solutions to the multiband Eliashberg equations that do not obey this symmetry.
Using the GPA allows us to make significant headway analytically. Although we cannot ascertain that the states we find are the only states satisfying the Eliashberg equations, the states we do find using the GPA clearly do.

Another property of the gap presented in Ref.\ \cite{Yuzbashyan2022a} is that $|\Phi(\omega_n)|$ is even in frequency. This is supported by the fact that renormalization effects seem to preclude odd-frequency superconductivity mediated by phonons \cite{Abrahams1993, Linder2019}, except in the presence of a magnetic field \cite{Aperis2015, Maeland2023}. Hence, assuming singlet pairing, the spatial symmetry of the gap must be even to satisfy the overall symmetry restrictions posed on the Cooper pairs \cite{Linder2019}. The only spatial symmetry satisfying Eq.\ \eqref{GPA} is $s$-wave symmetry, thus also satisfying the overall symmetry of the Cooper pairs. $s$-wave symmetry is the appropriate spatial symmetry in many systems, especially when phonons are mediating the interactions. If neither the phonon energies nor the electron-phonon coupling depends on momentum, it follows directly (also for multiband systems) from Eq.\ \eqref{eliash_eq_Phi} that $\Phi_i(k) = \Phi_i(\omega_n)$, since the right-hand side does not depend on $\mathbf{k}$, showing the $s$-wave symmetry of $\Phi_i$. $\Phi(k) = \Phi(\omega_n)$ also holds when dispersive phonons are mediating the interactions as long as the Fermi energy is the dominant energy in the system \cite{Eliashberg1960, Yuzbashyan2022a}.

The symmetry considerations above leave $\Phi_i(\omega_n)$ with almost the same characteristics as the gaps used in multiband BCS theory; spin-singlet $s$-wave gaps with global phases, but with an additional frequency dependence accounting for retardation effects. The similarity allows us to study multiband phenomena previously found in BCS theory, only now in the strong-coupling regime. One of these phenomena is the TRSB states arising in systems where the coupling between the gaps causes phase frustration that, in some cases, spontaneously breaks TRS. In BCS theory, the microscopic conditions under which TRSB occur have been outlined in detail \cite{Stanev2010, Wilson2013, Takahashi2014}. We will now spend the remainder of this section outlining the equivalent conditions in Eliashberg theory, and in the following two sections, we consider a specific model where they are satisfied.

An alternative way of understanding the criteria for TRSB to occur in multiband BCS theory is to consider the free energy. This approach is used in Ref.\ \cite{Johnsen2021}. Denoting the matrix containing the $k$-independent interactions between bands $i$ and $j$ as $V_{ij}^{\mathrm{BCS}}$ (where we will continuously add the superscript when needed to distinguish from the strong-coupling case), the phases of the gaps only appear in one term in the free energy, namely $\sum_{i,j}\bar{\Delta}_i^{\mathrm{BCS}} [V^{\mathrm{BCS}}]^{-1}_{ij}\Delta_j^{\mathrm{BCS}}$ \cite{Benfenati2023, Aase2023, Johnsen2021}. Splitting the BCS gaps into their amplitude and phase, this term becomes
$\sum_{i,j} a^{\mathrm{BCS}}_{ij}\cos\theta^{\mathrm{BCS}}_{ij}$ with $a^{\mathrm{BCS}}_{ij} = [V^{\mathrm{BCS}}]^{-1}_{ij} |\Delta_i^{\mathrm{BCS}}| |\Delta_j^{\mathrm{BCS}}|$. Note that $a^{\mathrm{BCS}}_{ij}$ is symmetric, so the distinct off-diagonal elements are $a^{\mathrm{BCS}}_{12}$, $a^{\mathrm{BCS}}_{13}$, and $a^{\mathrm{BCS}}_{23}$. Now, to have phase frustration in the three-band case, either one or all three of the off-diagonal elements in $a^{\mathrm{BCS}}_{ij}$ must be positive, akin to how a spin system on a triangular lattice is frustrated if all couplings are antiferromagnetic or if two are ferromagnetic and one is antiferromagnetic. The signs of the couplings are decided by $[V^{\mathrm{BCS}}]^{-1}_{ij}$.

For the phase frustration to give TRSB, the amplitude of the couplings must be such that the phases that minimize the free energy are not multiples of $\pi$. If this is the case, the ground state is twofold degenerate and related by time-reversal symmetry (complex conjugation). As the system enters one of these two distinct states, it breaks a $Z_2$ symmetry, which, in this case, corresponds to spontaneously breaking time-reversal symmetry.

By inserting Eq.\ \eqref{GPA} into the cross terms in the second term in Eq.\ \eqref{free_energy_with_momentum} and assuming $s$-wave spatial symmetry, one obtains
\begin{equation}
    \sum_{k, k'}\sum_{i,j}V^{-1}_{ij}(k-k') \bar{\Phi}_i(k) \Phi_j(k') = \sum_{i,j} a_{ij}\cos\theta_{ij} \label{phase_term},
\end{equation}
with
\begin{align}
    a_{ij} &= \sum_{k, k'}  Z_i(\omega_n)Z_j(\omega_m)|\Delta_i(\omega_n)||\Delta_j(\omega_m)| V_{ij}^{-1}(k-k') \label{a_ij},
\end{align}
where we used that $Z_i(\omega_n)$ is positive for all $\omega_n$ \cite{Zhang2022}.
In Eq.\ \eqref{a_ij}, we employed the Eliashberg fields, rather than the original HS fields, to elucidate the connection between $a_{ij}$ and $a_{ij}^{\mathrm{BCS}}$. The most apparent differences between the two are the double sum over $k, k'$ in $a_{ij}$ and the frequency dependency of the fields. Moreover, renormalization effects are accounted for in $Z_i(\omega_n)$ and $Z_j(\omega_m)$. However, the pivotal role in deciding the fate of TRSB is, akin to BCS theory, played by the inverse interaction. It determines the sign of $a_{ij}$ and, as mentioned previously, either one or all three $a_{ij}$ must be positive for TRSB to occur. 
Additionally, the mean-field values of the fields enter in $a_{ij}$. These are decided by the Eliashberg equations and thus by the interaction $V_{ij}(k-k')$. This creates a complex interplay between the interaction and its inverse, where both are needed to calculate $a_{ij}$. A numerical scheme for solving the Eliashberg equations that exploits this interplay is explained in Appendix \ref{app:num}, and we will return to it later.

\section{The microscopic model and its inverse interaction}\label{sec:micro}
Having kept the discussion rather general until now, we will in this section introduce the microscopic model that adheres to the action in Eq.\ \eqref{starting_point_Eliash}.
Moreover, we will calculate the inverse interaction of the model we employ and substantiate the claim we made in Eq.\ \eqref{loose_def_of_V_inv}.
As mentioned in relation to Eq.\ \eqref{starting_point_Eliash}, we assume that the system has three disjoint Fermi surfaces and enumerate the electrons belonging to each Fermi surface $i=1,2,3$. These Fermi surfaces may result from a band crossing the Fermi energy several times, resulting in multiple Fermi pockets, or from multiple bands, each crossing the Fermi energy once. Both cases are well-described by the multi-patch method \cite{Perali2001}, and in the following, we will use the terms band and pocket interchangeably.

We consider interactions between the electrons that are mediated by phonons. A complete derivation of the effective interaction in such systems using functional-integral formalism can be found in Ref.\ \cite{Protter2021}. Referring to this for details, we state the result, as the potential for one-band systems is $V(k) = -\beta^{-1}|g_{\mathbf{k}}|^2\mathcal{D}_0(k)$, where $g_{\mathbf{k}}$ is the electron-phonon coupling, depending, in general, on momentum, and $\mathcal{D}_0(k)$ is the phonon propagator, defined as
\begin{equation}
    \mathcal{D}_0(k) = \frac{2\omega_{\mathbf{k}}}{(i\varpi_n)^2 - \omega_\mathbf{k}^2},
    \label{D_0}
\end{equation}
where $\varpi_n$ is a bosonic Matsubara frequency and $\omega_\mathbf{k}$ is the phonon frequency at momentum $\mathbf{k}$.
We also inserted $\beta^{-1}$ in the interaction to ensure that it carries units of energy squared in real and momentum space with our Fourier transform conventions. 

To obtain a multiband interaction, we will make some changes to the interaction above. Firstly, to account for the different geometries of the Fermi pockets and how the electrons belonging to a pocket couple to phonons, we take the electron-phonon coupling to depend on which bands are involved in the scattering.
For example, one can imagine that pockets $i=1$ and $i=2$ are spherical but the radius of the $i=1$ Fermi pocket is larger. The average momentum transferred is then larger in scattering processes for electrons belonging to pocket $i=1$. So when we approximate the electron-coupling to be momentum independent, we account for this by including band indices in $|g_{\mathbf{k}}|^2\rightarrow |g_{ij}|^2$. Note that $|g_{ij}|$ is symmetric.

We make a second simplification in assuming that all phonons have frequency $\omega_{\mathrm{E}}$, similar to an Einstein solid. As we will see later, this is not an essential simplification, but we note that it has been used in similar works on strong coupling \cite{Yuzbashyan2022a, Zhang2022, Dalal2023}. With it, $\omega_\mathrm{E}$ replaces $\omega_\mathbf{q}$ in Eq.\ \eqref{D_0} and the multiband interaction becomes
\begin{equation}
    V_{ij}(\omega_n - \omega_m) = \frac{1}{\beta} |g_{ij}|^2 \frac{2 \omega_{\mathrm{E}}}{(\omega_n - \omega_m)^2 + \omega_{\mathrm{E}}^2},
    \label{V_def}
\end{equation}
which is the interaction we will use in the following.

\begin{figure}
    \centering
    \includegraphics[width=8.6cm,height=6.1cm,keepaspectratio]{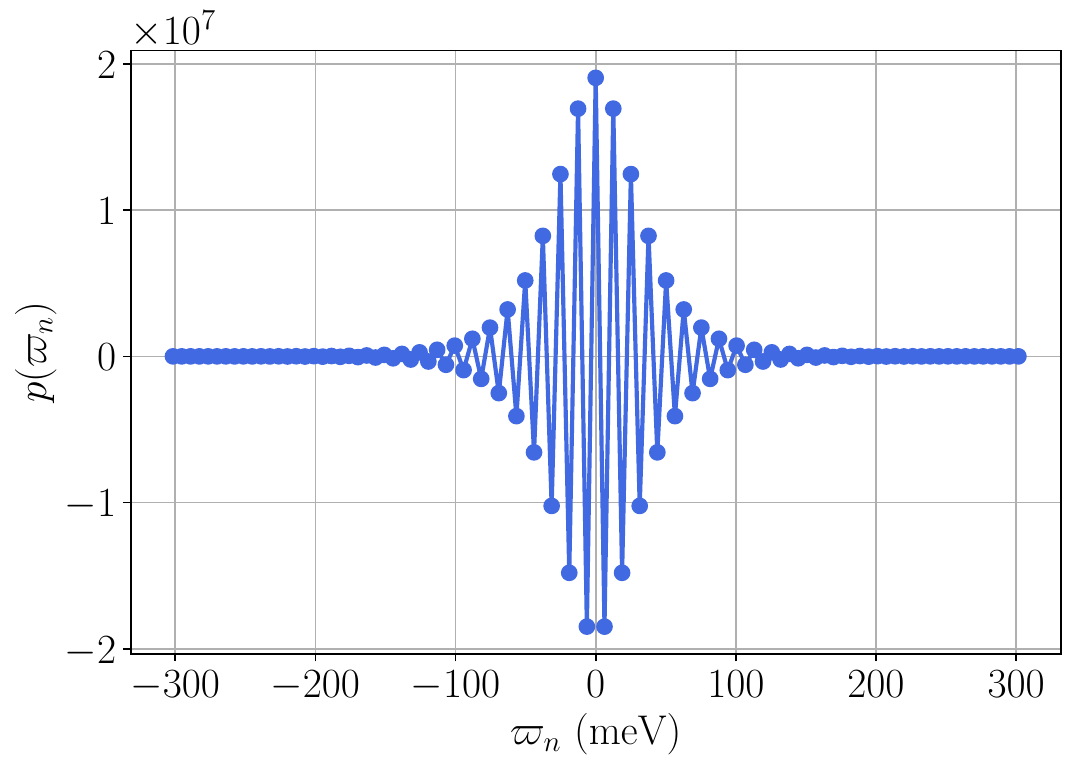}
    \caption{The frequency-dependent part of the inverse interaction $p(\varpi_n)$ as a function of the bosonic Matsubara frequencies $\varpi_n$. The Einstein frequency is $\omega_{\mathrm{E}} = 40$ meV and the temperature is $T=1$ meV.}
    \label{fig:p_func}
\end{figure}

It is convenient to split the interaction into a band part and a frequency part. To denote the former, we introduce the symmetric matrix $B_{ij} = |g_{ij}|^2$, such that all elements in $B$ are nonnegative. The frequency dependence comes from the phonon propagator $\mathcal{D}_0(\omega_n-\omega_m =\varpi_n)$, always evaluated at the difference between two fermionic Matsubara frequencies which is itself a bosonic one.
We now carry out the steps required to calculate the inverse interaction, as outlined in Eq.\ \eqref{loose_def_of_V_inv}.
The temporal Fourier transform of $V_{ij}(\varpi_n)$ is straightforward to calculate, see e.g.\ \cite{Atland2010}
\begin{equation}
    V_{ij}(\tau) = \frac{B_{ij}}{\beta}\sum_{\varpi_n} \frac{2 \omega_{\mathrm{E}} \mathrm{e}^{-i\varpi_n\tau}}{\varpi_n^2+ \omega_{\mathrm{E}}^2} = B_{ij} \frac{\mathrm{e}^{\omega_{\mathrm{E}}(\beta - \tau)} + \mathrm{e}^{\omega_{\mathrm{E}}\tau}}{\mathrm{e}^{\beta\omega_{\mathrm{E}}} - 1}. \label{V_ij_tau}
\end{equation}
The matrix inverse of Eq.\ \eqref{V_ij_tau} is simply
\begin{equation}
    V^{-1}_{ij}(\tau) = B_{ij}^{-1} \frac{\mathrm{e}^{\beta\omega_{\mathrm{E}}} - 1}{\mathrm{e}^{\omega_{\mathrm{E}}(\beta - \tau)} + \mathrm{e}^{\omega_{\mathrm{E}}\tau}}.
\end{equation}
To complete the derivation of the inverse interaction, we return to Fourier space
\begin{align}
    V^{-1}_{ij}(\varpi_n) &= B_{ij}^{-1} \frac{\mathrm{e}^{\beta\omega_{\mathrm{E}}} - 1}{\beta}\int_0^\beta \mathrm{d}\tau \frac{\mathrm{e}^{(i\varpi_n +\omega_{\mathrm{E}})\tau}}{\mathrm{e}^{\beta\omega_{\mathrm{E}}} + \mathrm{e}^{2\omega_{\mathrm{E}}\tau}} \nonumber \\
    &= B_{ij}^{-1}p(\varpi_n) \label{V_inv_omega}, 
\end{align}
with
\begin{align}
    p(\varpi_n) &= \frac{\mathrm{e}^{\beta \omega_E}-1}{\beta}\bigg[ \frac{F(1, \frac{i\varpi_n + \omega_{\mathrm{E}}}{2\omega_{\mathrm{E}}}; \frac{i\varpi_n + 3\omega_{\mathrm{E}}}{2\omega_{\mathrm{E}}}; -\mathrm{e}^{-\beta\omega_{\mathrm{E}}})}{i\varpi_n + \omega_{\mathrm{E}}} \nonumber \\
    &- \frac{F(1, \frac{i\varpi_n + \omega_{\mathrm{E}}}{2\omega_{\mathrm{E}}}; \frac{i\varpi_n + 3\omega_{\mathrm{E}}}{2\omega_{\mathrm{E}}}; -\mathrm{e}^{-\beta\omega_{\mathrm{E}}})}{\mathrm{e}^{\beta\omega_{\mathrm{E}}}(i\varpi_n + \omega_{\mathrm{E}})} \bigg] \label{p_def}.
\end{align}
$F(a, b; c; z)$ is the hypergeometric function \cite{Gradshteyn1980}. In Fig.\ \ref{fig:p_func}, $p(\varpi_n)$ is plotted as a function of $\varpi_n$ with $T=1$ meV and $\omega_{\mathrm{E}} = 40$ meV. $p(\varpi_n)$ is real, even in $\varpi_n$, and changes sign between every adjacent $\varpi_n$.
For the expression in Eq.\ \eqref{V_inv_omega} to be the inverse interaction it has to satisfy the relation in Eq.\ \eqref{inverse_relation}. This is straightforward to demonstrate when expressing $V_{ij}(\varpi_n)$ and $V^{-1}_{ij}(\varpi_n)$ in their imaginary time representation $v(\tau) = (\mathrm{e}^{\beta\omega_{\mathrm{E}}} - 1)(\mathrm{e}^{\omega_{\mathrm{E}}(\beta - \tau)} + \mathrm{e}^{\omega_{\mathrm{E}}\tau})^{-1}$ and $\frac{1}{v(\tau)}$, respectively, 
\begin{align}
    &\sum_{\omega_n} \sum_l V_{il}^{-1}(\omega_{n_1} - \omega_n) V_{lj}(\omega_n-\omega_{n_2}) \nonumber \\
    &= \sum_l B^{-1}_{il}B_{lj}\frac{1}{\beta}\int_0^\beta \mathrm{d}\tau_1\mathrm{d}\tau_2 \frac{v(\tau_1)}{v(\tau_2)} \mathrm{e}^{-i\omega_{n_1}\tau_1}\mathrm{e}^{i\omega_{n_2}\tau_2}\nonumber \\
    &\times \frac{1}{\beta}\sum_{\omega_n} \mathrm{e}^{i\omega_n(\tau_1 -\tau_2)} \nonumber \\
    &= \delta_{ij} \frac{1}{\beta}\int_0^\beta \mathrm{d}\tau_1\mathrm{d}\tau_2\frac{v(\tau_1)}{v(\tau_2)} \mathrm{e}^{-i\omega_{n_1}\tau_1}\mathrm{e}^{i\omega_{n_2}\tau_2} \delta(\tau_1-\tau_2) \nonumber \\
    &= \delta_{ij} \delta_{\omega_{n_1}\omega_{n_2}}.
\end{align}
Now, with the inverse interaction at hand, we reiterate one of the main messages from Sec.\ \ref{sec:GPA}: While it is the interaction that determines the solutions to the Eliashberg equations (requiring numerics), the conditions under which TRSB may occur can be gleaned directly from the inverse interaction.
In particular, the signs of $a_{ij}$ are the same as $B_{ij}^{-1}$ in all the systems we have considered, meaning that the double frequency sum in $a_{ij}$,
\begin{align}
    &a_{ij}= \nonumber \\
    &B_{ij}^{-1}  \sum_{\omega_n, \omega_m} Z_i(\omega_n)Z_j(\omega_m)|\Delta_i(\omega_n)||\Delta_j(\omega_m)|p(\omega_n-\omega_m) \label{a_ij_freq}
\end{align}
converges to a positive number. We again note that $a_{ij}$ is symmetric. Thus, a necessary condition for spontaneously breaking TRS is that one or all three of the off-diagonal elements in $B_{ij}^{-1}$ are positive. If not, there is no phase frustration and consequently no TRSB.

An intriguing way of generalizing the interaction considered here is outlined in Appendix B in Ref.\ \cite{Yuzbashyan2022a}. For dispersive phonons where the frequency $\omega_{q}$ is only dependent on the magnitude of $\mathbf{q}$, denoted $q$, one finds spatially isotropic solutions to the Eliashberg equations by exploiting the symmetries of the Fermi surface, which in Ref.\ \cite{Yuzbashyan2022a} is taken to be spherical. The momentum dependence then renormalizes the interaction in the form
\begin{equation}
    \Tilde{V}(\varpi_n) = \frac{|g|^2}{2k_{\mathrm{F}}\beta} \int_0^{2k_\mathrm{F}} \mathrm{d}q \frac{f(q)\omega_q}{\varpi_n^2 + \omega_q^2},
    \label{V_tilde}
\end{equation}
where $k_{\mathrm{F}}$ is the Fermi momentum, $\Tilde{V}(\varpi_n)$ is the renormalized potential, and $f(q)$ is a dimensionless function determined by the dimensionality of the system. To generalize Eq.\ \eqref{V_tilde} to a multiband system, one must consider the different geometries of the Fermi surfaces, which would affect both $f(q)$ and the integration limits.  $|g|^2$ would be replaced by $B_{ij}$. The steps of calculating $\Tilde{V}^{-1}_{ij}(\varpi_n)$ are then the same as the ones leading up to Eq.\ \eqref{V_inv_omega}. While $V_{ij}(\varpi_n)$ has the advantage of only having a few free parameters in $\omega_{\mathrm{E}}$ and $g_{ij}$, $\Tilde{V}_{ij}(\varpi_n)$ would be able to utilize first-principles calculations for the electron band structure and associated Fermi surface, the full phonon spectrum, and the electron-phonon coupling \cite{Giustino2017}, thus facilitating a more effective search for candidate materials where TRSB might arise. We leave this for future work, and keep to the simpler interaction in Eq.\ \eqref{V_def} going forward. 

To close this section, we simplify the Eliashberg equations and the free energy in Sec.\ \ref{sec:funcint}. The momentum sums can be carried out analytically since the Eliashberg fields only depend on frequency. As the momentum dependence is only in $\xi_i(\mathbf{k})$, the sums can be transformed into energy integrals in standard fashion \cite{Marsiglio2020}. Approximating the density of states of each band by its value at the Fermi surface, denoted $N_{\mathrm{F}, i}$, the energy integrals may be carried out. In the case of $\chi_i(\omega_n)$ in Eq.\ \eqref{eliash_eq_for_chi}, the energy integral, and thus also $\chi_i(\omega_n)$ itself, is zero. This simplifies the expression for $\Theta_i(k)$ in Eq.\ \eqref{determinant}, which contains the only energy dependence in the remaining two Eliashberg equations. After integrating over energy, these become
\begin{align}
    Z_i(\omega_n) &= 1 + \frac{\pi}{\omega_n}\sum_{\omega_m, j}\frac{V_{ij}(\omega_n-\omega_m)N_{\mathrm{F}, j} \omega_m}{\sqrt{\omega_m^2 + |\Delta_j(\omega_m)|^2}} \label{finished_eliash_eq_for_Z}\\
    \Delta_i(\omega_n) &= \frac{\pi}{Z_i(\omega_n)}\sum_{\omega_m, j}   \frac{V_{ij}(\omega_n-\omega_m)N_{\mathrm{F}, j}\Delta_j(\omega_m)}{\sqrt{\omega_m^2 + |\Delta_j(\omega_m)|^2}}.\label{finished_eliash_eq_for_Delta}
\end{align}
Similarly, following the procedure of Ref.\ \cite{Yuzbashyan2022a}, by sending the Fermi energy to infinity and discarding constant terms, we integrate the logarithm in the free energy in Eq.\ \eqref{free_energy_with_momentum} and obtain
\begin{align}
    \beta f&= -2\pi\sum_{\omega_n}\sum_i N_{\mathrm{F, i}}Z_i(\omega_n)\sqrt{\omega_n^2 +|\Delta_i(\omega_n)|^2}  \nonumber\\
    &+ \sum_{\omega_n,\omega_m}\sum_{i,j}V^{-1}_{ij}(\omega_n-\omega_m) \nonumber \\
    &\times[\bar{\Phi}_i(\omega_n) \Phi_j(\omega_m) + \Sigma_i(\omega_n)\Sigma_j(\omega_m)]. \label{free_energy_with_only_freq}
\end{align}
Here, we introduced the free energy density $f\equiv F/N$, where $N$ is the number of unit cells.

The Eliashberg equations above coincide with the multiband Eliashberg equations used in previous works \cite{Nicol2005, Ummarino2009}. However, equipped with the global phase ansatz and the expression for $a_{ij}$ in Eq.\ \eqref{a_ij_freq}, we are in a position to investigate possible TRSB states not seen before in strong-coupling models.

\section{Results}\label{sec:results}
In this section, we will showcase a system that spontaneously breaks TRS in strong coupling, manifesting in nonanalytical behavior in both the phases and amplitudes of the gaps due to a $Z_2$-breaking Ising phase transition at $T=T^*$.
We will discuss other systems that might exhibit TRSB later.

The system under consideration here consists of three coupled bands. Two of the bands have strong intraband interactions of different magnitudes. The remaining interactions are weaker, $g_{11}\approx g_{22}\gg g_{ij}$. In units of $g=85$ meV, the electron-phonon couplings between the different bands are $g_{22} = 1.4 g$, $g_{33} = 0.005 g$, $g_{12} = 0.0001 g$, $g_{13} = g_{23} = 0.003g$, while the intraband coupling in the first band $g_{11}$ can vary but is, in general, of the same order as $g$. For simplicity, we set the density of states at the Fermi level equal for all three bands $N_{\mathrm{F}, i} = 0.5 \cdot 10^{-2}$ meV$^{-1}$, and the Einstein frequency and temperature are $20$ meV and $1$ meV, respectively. With these values for $g_{ij}$, one obtains $B_{12}^{-1} \approx 2 \cdot10^{-10}$, $B_{13}^{-1} \approx -5 \cdot10^{-5}$ and, $B_{23}^{-1} \approx -3 \cdot10^{-5}$, all in units of meV$^{-2}$.

Before showing results from this system, it is worth considering the type of physical systems where TRSB may be realized. Firstly, we consider two intraband couplings approximately equal to $g=85$ meV, putting the system into the strong-coupling regime. The system is also one where the interband electron-phonon coupling is severely suppressed. This is the case in iron pnictides \cite{Mazin2009, Ummarino2009, Hirschfeld2011}, some of which are believed to be strong-coupling superconductors \cite{Jo2009, Xing2018}. Lastly, a possible realization of the small intraband interaction in the third band is the following: As explained in Sec.\ \ref{sec:micro}, the electron-phonon coupling depends on the momentum transferred in the scattering. Moreover, for the case of Einstein phonons, $g_{\mathbf{k}} \propto |\mathbf{k}|$ \cite{Bruus2004}. Therefore, if bands 1 and 2 (band 3) have large (small) Fermi surfaces, the average momentum transfer, and thus also the electron-phonon coupling, will be bigger (smaller). This can be finetuned by tuning the chemical potential such that both the Fermi surface and the electron-phonon coupling of the third band become small. We discuss our specific choice of parameters in more detail in Appendix \ref{app:parameters}.

\begin{figure*}[t]
    \centering
    \includegraphics[width=18cm,height=9.cm,keepaspectratio]{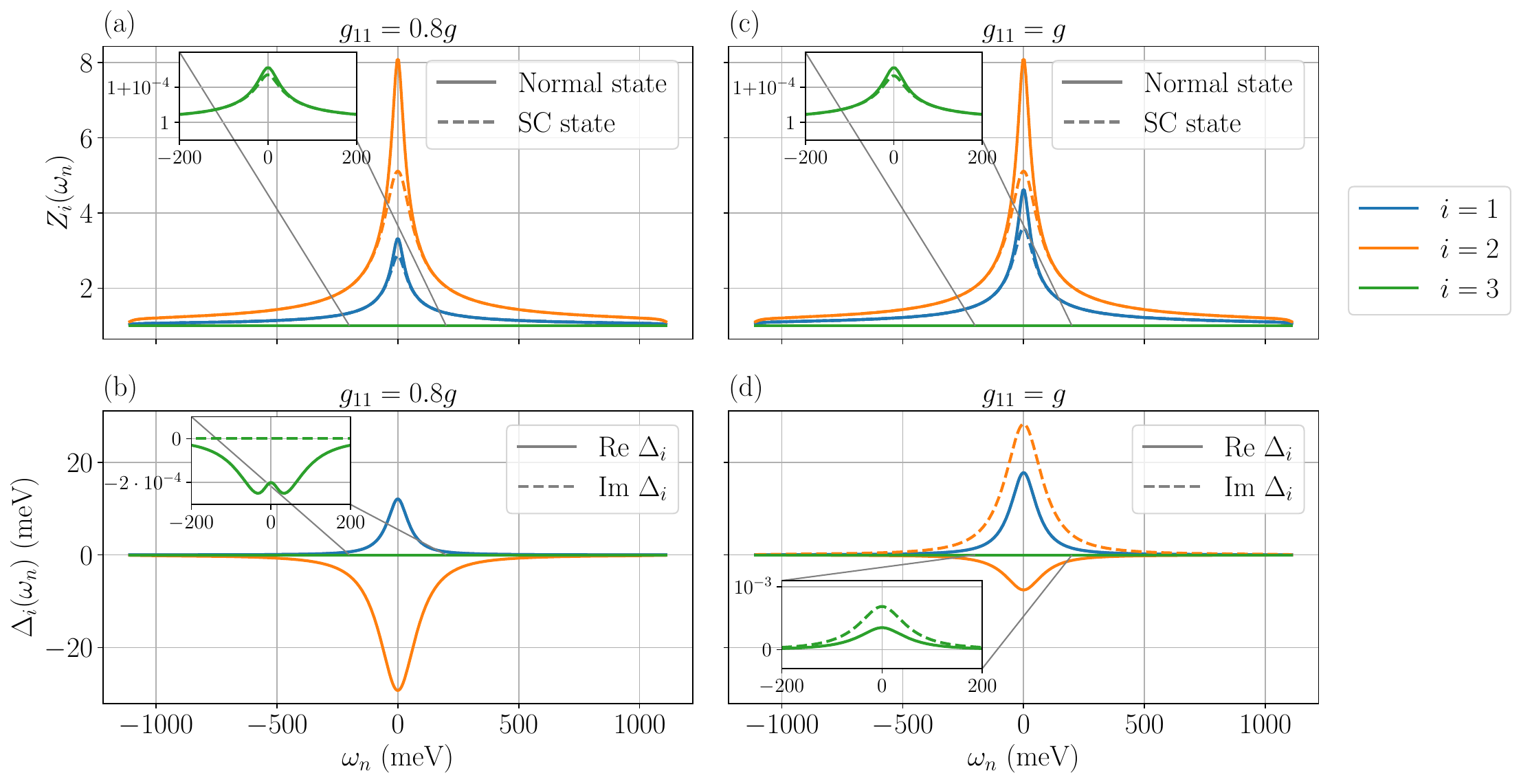}
    \caption{The solutions to the three-band Eliashberg equations for the inverse quasiparticle residue $Z_i(\omega_n)$ and the superconducting gap $\Delta_i(\omega_n)$ with band index $i$. In the two left panels, the intraband electron-phonon coupling in the first band is $g_{11}=0.8g$ with $g$ = 85 meV, and it is $g_{11}=g$ in the right panels. Otherwise, the system parameters are the same, where the interaction parameters are constant and equal to $g_{22} = 1.4 g$, $g_{33} = 0.005 g$, $g_{12} = 0.0001 g$, $g_{13} = g_{23} = 0.003g$ and the other system parameters are $T=1$~meV, $\omega_{\mathrm{E}} =20$ meV, and $N_{\mathrm{F}, i} = 0.5 \cdot 10^{-2}$ meV$^{-1}$. In panels (a) and (c), the solutions are found in both the normal (solid lines) and superconducting (dashed lines) state, while in panels (b) and (d), the real (imaginary) part of the gaps is plotted with solid (dashed) lines. The insets in all panels show the fields associated with band index $i=3$.}
    \label{fig:sols_with_freq_dep}
\end{figure*}

It is instructive to see how and why this causes TRSB, which follows directly from the expression for $a_{ij}$ in Eq. \eqref{a_ij_freq}. Since the signs of $a_{ij}$ are determined by $B_{ij}^{-1}$, the system will experience some phase frustration, as discussed in Sec.\ \ref{sec:micro}. Moreover, because $g_{11}\approx g_{22} \gg g_{33}$, it follows that $\Delta_1 \approx \Delta_2 \gg \Delta_3$ below $T_{\mathrm{c}}$.
Therefore, again from Eq.\ \eqref{a_ij_freq}, we expect similar magnitudes of $a_{12}$, $a_{13}$, and $a_{23}$, thus possibly setting the stage for the phase frustration to result in TRSB.

Following the numerical scheme in Appendix \ref{app:num}, we solve the multiband Eliashberg equations in Eqs.\  \eqref{finished_eliash_eq_for_Z} and \eqref{finished_eliash_eq_for_Delta} with the parameters above. The results are illustrated in Fig.\ \ref{fig:sols_with_freq_dep}, where we have studied two similar systems with different $g_{11}$.
Because $g_{33}\ll g_{11}, g_{22}$, the fields of the third bands are significantly smaller than their counterparts in bands 1 and 2. For this reason, we have included insets to show the behavior of $Z_3(\omega_n)$ and $\Delta_3(\omega_n)$ more clearly.
All fields are even in $\omega_n$. $Z_i(\omega_n)$ decays as a function of $|\omega_n|$ in all bands, both in the case of $g_{11}=0.8g$ and $g_{11} = g$, as shown in Figs.\ \ref{fig:sols_with_freq_dep} (a) and (c), respectively.
As $g_{11}$ grows from Fig.\ \ref{fig:sols_with_freq_dep} (a) to (c), so does $Z_1(\omega_n)$, which is explained by stronger renormalization effects, induced by the increase of the intraband electron-phonon coupling. $Z_2(\omega_n)$ and $Z_3(\omega_n)$ remain largely unaffected by changing $g_{11}$. Moreover, we observe that $Z_i(\omega_n)$ have larger values in the normal state, which is explained by the factor $\sqrt{\omega_m^2 + |\Delta_j(\omega_n)|^2}$ in the denominator in Eq.\ \eqref{finished_eliash_eq_for_Z}.

In Fig.\ \ref{fig:sols_with_freq_dep} (b), $\Delta_i(\omega_n)$ is plotted as a function of frequency with $g_{11}=0.8g$. All gaps are purely real, where the phase differences $\theta_{12}$ and $\theta_{13}$ are $\pi$.
We note that the realvaluedness of all gaps is not guaranteed in general since the global $U(1)$ symmetry can only be chosen such that one of the gaps is real and positive. We take this gap to be $\Delta_1(\omega_n)$ in the following.
Moreover, because $g_{22}>g_{11}$, it follows that $|\Delta_2|>|\Delta_1|$. Both gaps display the same monotonic decreasing behavior with $|\omega_n|$. This is not the case for the smallest gap, as $|\Delta_3(\omega_n)|$ increases slightly with $|\omega_n|$, quickly reaching its peak before it becomes monotonically decaying as well.
To elucidate the behavior of the third gap, it is useful to consider the mechanisms sustaining the different gaps. The bands with strong electron-phonon coupling are superconducting in their own right; both $\Delta_1(\omega_n)$ and $\Delta_2(\omega_n)$ survive if the interband couplings vanish. Their amplitude would barely change. For $\Delta_3(\omega_n)$, however, $g_{33}$ is far too weak to drive the superconductivity by itself. Instead, $\Delta_3(\omega_n)$ is sustained by the other two gaps, from $j=1$ and $j=2$ on the right-hand side in Eq.\ \eqref{finished_eliash_eq_for_Delta}. The mechanism of a weak gap being supported by larger ones has long been recognized in multiband superconductors \cite{Suhl1959}.
Since $|\Delta_2(\omega_n)|>|\Delta_1(\omega_n)|$ and $g_{13}=g_{23}$, the $j=2$ term is larger in magnitude than the $j=1$ term, causing $\Delta_3(\omega_n)$ to have the same sign as $\Delta_2(\omega_n)$.

The gaps, unlike $Z_i(\omega_n)$, exhibit a more abrupt change when $g_{11}$ increases in Fig.\ \ref{fig:sols_with_freq_dep} (d). The most apparent effect is that $\theta_{ij}$ are no longer multiples of $\pi$, as $\Delta_2(\omega_n)$ and $\Delta_3(\omega_n)$ both have real and imaginary parts, while $\Delta_1(\omega_n)$ is still purely real. Therefore the system in Fig.\ \ref{fig:sols_with_freq_dep} (d) spontaneously breaks TRS since the state is not invariant under complex conjugation.
The slight increase in $g_{11}$ also affects the magnitudes of the lowest-frequency gaps $|\Delta_i(\omega_{n=0})| \equiv |\Delta_i(0)|$. As expected, increasing $g_{11}$ causes $|\Delta_1(0)|$ to be larger, while $|\Delta_2(0)|$ remains the same. It is more surprising that the small change in $g_{11}$ also increases $|\Delta_3(0)|$ by a factor of 4. This can be understood from the fact that $\Delta_1(\omega_n)$ and $\Delta_2(\omega_n)$ are now less antagonistic in their support of $\Delta_3(\omega_n)$. In Fig.\ \ref{fig:sols_with_freq_dep} (b), $\theta_{12}$ was $\pi$, and thus the contributions to $\Delta_3(\omega_n)$ in Eq.\ \eqref{finished_eliash_eq_for_Delta} from $j=1$ and $j=2$ had opposite sign, as discussed previously. This is not the case in the system in Fig.\ \ref{fig:sols_with_freq_dep} (d), where $\Delta_1(\omega_n)$ ($\Delta_2(\omega_n)$) can boost the real (imaginary) part of $\Delta_3(\omega_n)$ without being counteracted to the same degree as before.

The antagonistic mechanism can also explain the small bump in $\Delta_3(\omega_n)$ in Fig.\ \ref{fig:sols_with_freq_dep} (b), as $\Delta_1(\omega_n)$ decreases faster than $\Delta_2(\omega_n)$, resulting in a small increase in $\Delta_3(\omega_n)$ at low frequencies even though both $\Delta_1(\omega_n)$ and $\Delta_2(\omega_n)$ decay with $|\omega_n|$. Conversely, in the system in Fig.\ \ref{fig:sols_with_freq_dep} (d), the magnitude of $|\Delta_3(\omega_n)|$ decays with $|\omega_n|$ such that all three gaps decay monotonically with $|\omega_n|$.

\begin{figure}
    \centering
    \includegraphics[width=8.6cm,height=15.6cm,keepaspectratio]{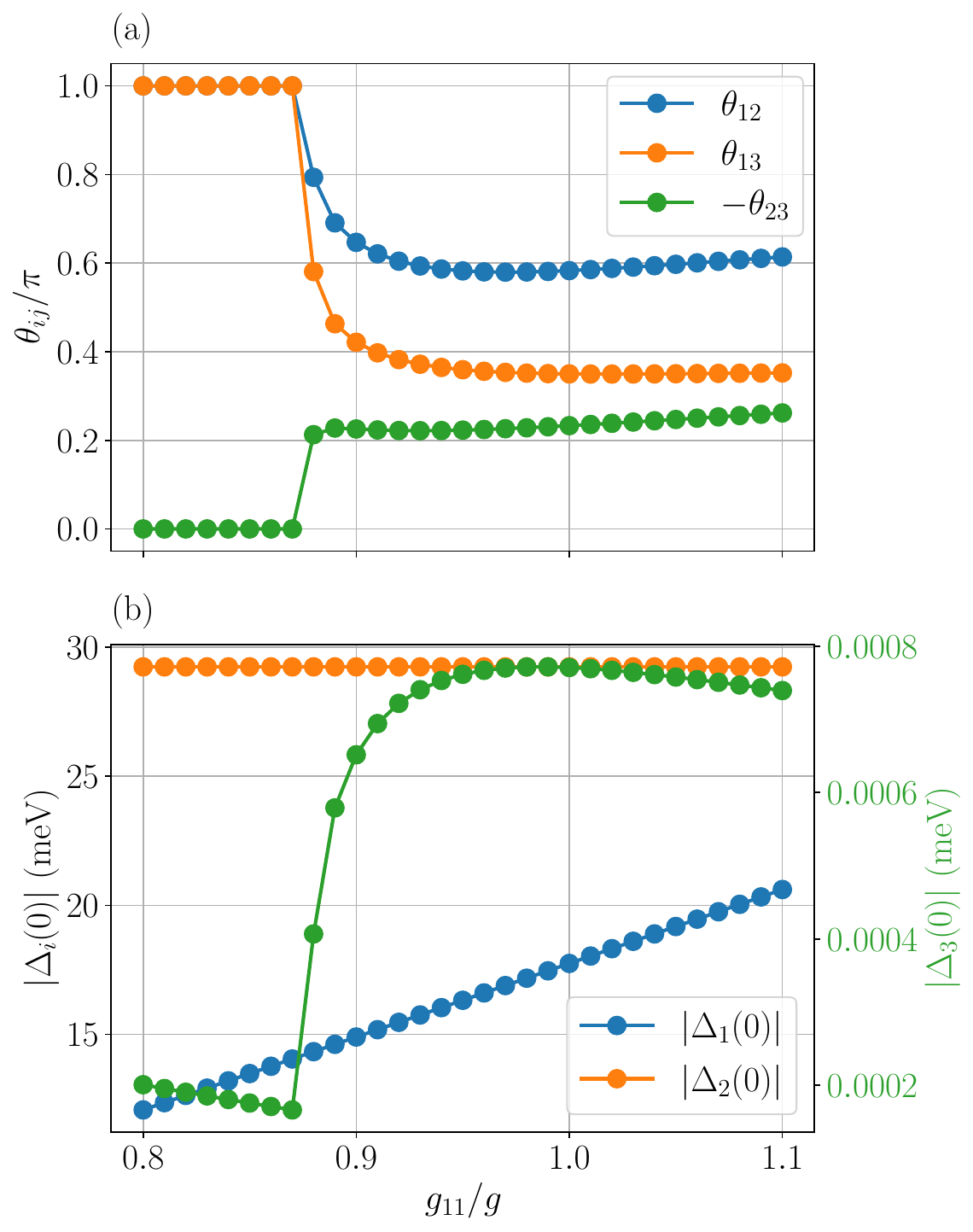}
    \caption{Results from solving the Eliashberg equations as a function of the intraband strength in the first band $g_{11}$ divided by $g = 85$ meV. In panel (a), the global phase differences $\theta_{ij}$ are plotted as a function of $g_{11}/g$. In panel (b), the lowest frequency gap in each band $\Delta_i(0)$ is plotted as a function of $g_{11}/g$. Note the different energy scales for $\Delta_1$ and $\Delta_2$ (left axis) and $\Delta_3$ (right axis). The other parameters are the same as in Fig.\ \ref{fig:sols_with_freq_dep}.}
    \label{fig:TRSB_as_func_of_g_11}
\end{figure}

To elucidate the transition between the TRS and TRSB states, we have solved the Eliashberg equations as a function of $g_{11}$, where the results are shown in Fig.\ \ref{fig:TRSB_as_func_of_g_11}. With the exception of $g_{11}$, we employ the same system parameters as in Fig.\ \ref{fig:sols_with_freq_dep}. In Fig.\ \ref{fig:TRSB_as_func_of_g_11} (b), $|\Delta_i(0)|$ is plotted as a function of $g_{11}/g$. For $|\Delta_1(0)|$ ($|\Delta_2(0)|$), we observe what we expect, namely that it increases (remains the same) with $g_{11}$. $|\Delta_3(0)|$, however, exhibit nonanalytical behavior at $g_{11} = 0.87g \equiv g_{\mathrm{c}}$. It is instructive to view these results in light of Fig.\ \ref{fig:TRSB_as_func_of_g_11} (a), which showcases $\theta_{ij}$ for the same values of $g_{11}$. For $g_{11}\leq g_{\mathrm{c}}$, $\theta_{12} =\pi$ and thus, due to the antagonistic mechanisms sustaining $\Delta_3(\omega_n)$ mentioned previously, $|\Delta_3|$ decreases when $|\Delta_1(0)|$ increases. This changes when TRS is broken at $g_{11}=g_{\mathrm{c}}$ such that $\theta_{12}$ falls off when $g_{11}>g_{\mathrm{c}}$, allowing $\Delta_1(\omega_n)$ and $\Delta_2(\omega_n)$ to boost the real and imaginary parts of $\Delta_3(\omega_n)$, respectively. The maximum of $\Delta_3(0)$ thus coincides with the minimum of $\theta_{12}$ at $g_{11}\approx 0.98g$, where $\Delta_1(\omega_n)$ and $\Delta_2(\omega_n)$ are at their most cooperative in sustaining $\Delta_3(\omega_n)$.

\begin{figure}
    \centering
    \includegraphics[width=8.6cm,height=10.1cm,keepaspectratio]{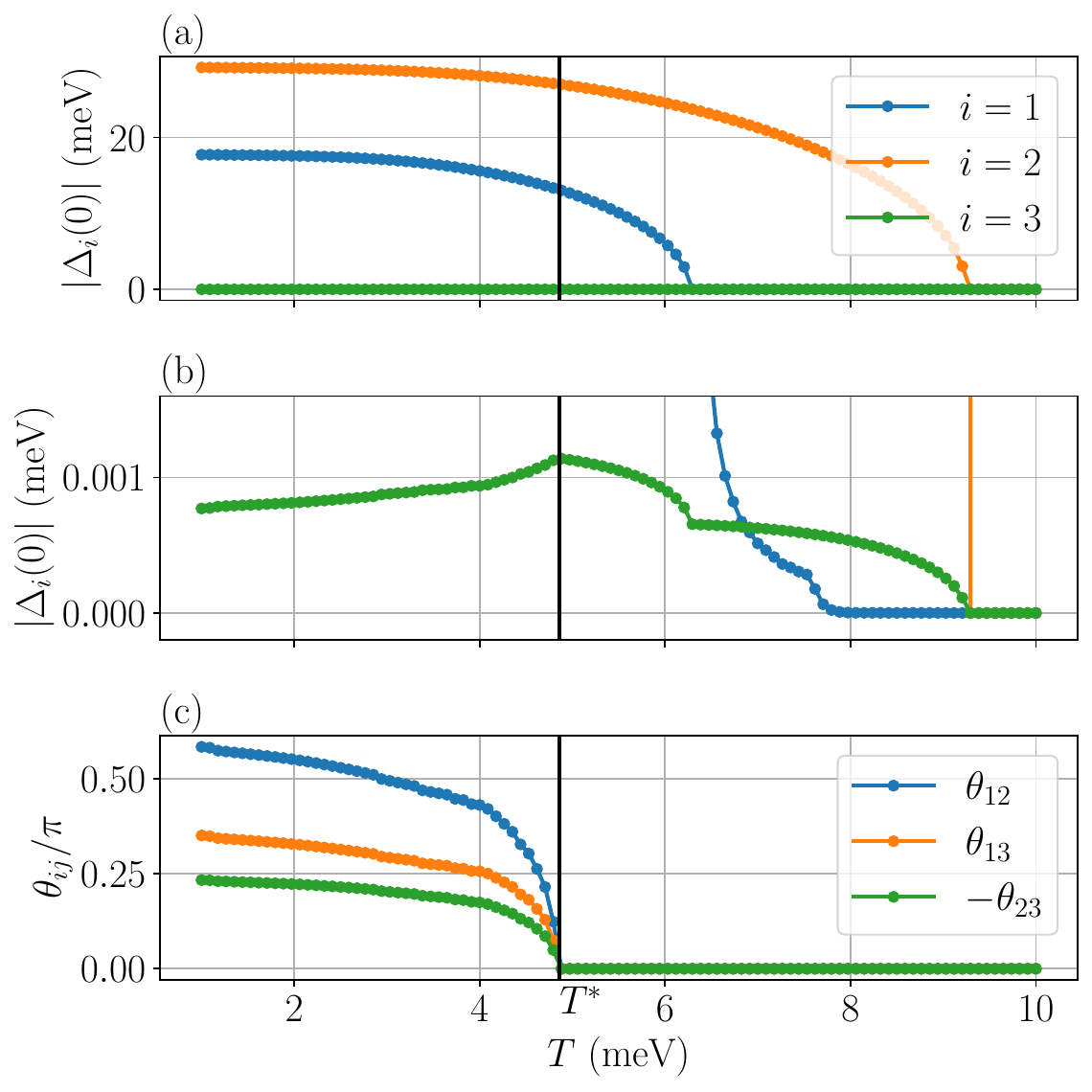}
    \caption{In panels (a) and (b), the lowest frequency gap amplitude $|\Delta_i(0)|$ in each band $i$ is plotted as a function of temperature $T$, where the panels have different scales on the $y$-axis to capture both the low and high-energy behavior. The phase differences between the bands $\theta_{ij}$ are plotted in panel (c). The system parameters are the same as in Fig.\ \ref{fig:sols_with_freq_dep} with $g_{11} = g$ = 85 meV. TRS is broken at $T<T^* = 4.85$ meV, and black vertical lines at $T=T^*$ are included to help guide the eye.}
    \label{fig:temperature_dependence}
\end{figure}

We emphasize that while $\Delta_3(\omega_n)$ is orders of magnitudes smaller than $\Delta_1(\omega_n)$ and $\Delta_2(\omega_n)$, it still plays a pivotal role in breaking TRS. If $\Delta_3(\omega_n)=0$, it follows immediately from Eq.\ \eqref{a_ij_freq} that $a_{12}$ is the only surviving term, such that $\theta_{12}$ is always zero or $\pi$. In the case of the latter, the resulting state is $s_\pm$. To obtain the $s+is$ state, all three $a_{ij}$ must be nonzero, in turn requiring all gaps to be nonzero.
In this sense, the TRSB will be robust, as $\theta_{ij}$ obtains values far away from zero or $\pi$ even though the third gap is small.
Moreover we expect the phenomena associated with TRSB to persist even if the third gap is small for the following reasons: 
In three-band microscopic models with TRSB, one can map such models to an effective Ginzburg-Landau functional with only two gaps where the TRSB is explicitly imposed \cite{Garaud2017, Lin2016}. The mapping does not affect the emergent physics such as spontaneous supercurrents near defects \cite{Lin2016} and topological invariants \cite{Garaud2017}. For the system considered here, the natural choice would then be to include the two large gaps into such an effective TRSB two-band model, but keeping in mind that the imposed TRSB is a consequence of the smaller gap. We leave this effective model for future consideration and keep exploring the properties of the ground state in the microscopic model going forward.

While Figs.\ \ref{fig:sols_with_freq_dep} and \ref{fig:TRSB_as_func_of_g_11} explain many of the mechanisms at play in causing TRSB, it is also instructive to consider the effects of externally controllable parameters. In Figs.\ \ref{fig:temperature_dependence} (a) and (b), we plot $|\Delta_i(0)|$ as a function of $T$ at two different energy scales to capture the behavior of both the large and small gaps. In the same temperature regime, we plot $\theta_{ij}$ in Fig.\ \ref{fig:temperature_dependence} (c). The critical temperature of the system is $T_\mathrm{c} \approx 9.2$ meV, where all gaps are zero for larger temperatures. $|\Delta_1(0)|$ onsets to larger values at $T_{1} \approx 6.2$ meV, but as shown in Fig.\ \ref{fig:temperature_dependence} (b), it retains nonzero values for larger values of $T$ as well. This is, again, a result of a stronger gap sustaining smaller ones through interband coupling \cite{Suhl1959}. Another onset is shown in Fig.\ \ref{fig:temperature_dependence} (c), namely that of $\theta_{ij}$ at $T=T^*=4.85$ meV. $T^*$ is thus the temperature at which TRS is spontaneously broken. $T^*$ and $T_1$ are the two temperatures lower than $T_\mathrm{c}$ where $|\Delta_3(0)|$ is nonanalytic. In the case of the latter, the sudden decrease in $|\Delta_3(0)|$ is due to the drop of $\Delta_1(\omega_n)$ at $T_1$, meaning that $\Delta_3(\omega_n)$ loses one of the sources maintaining it.

\begin{figure}
    \centering
    \includegraphics[width=8.6cm,height=8.6cm,keepaspectratio]{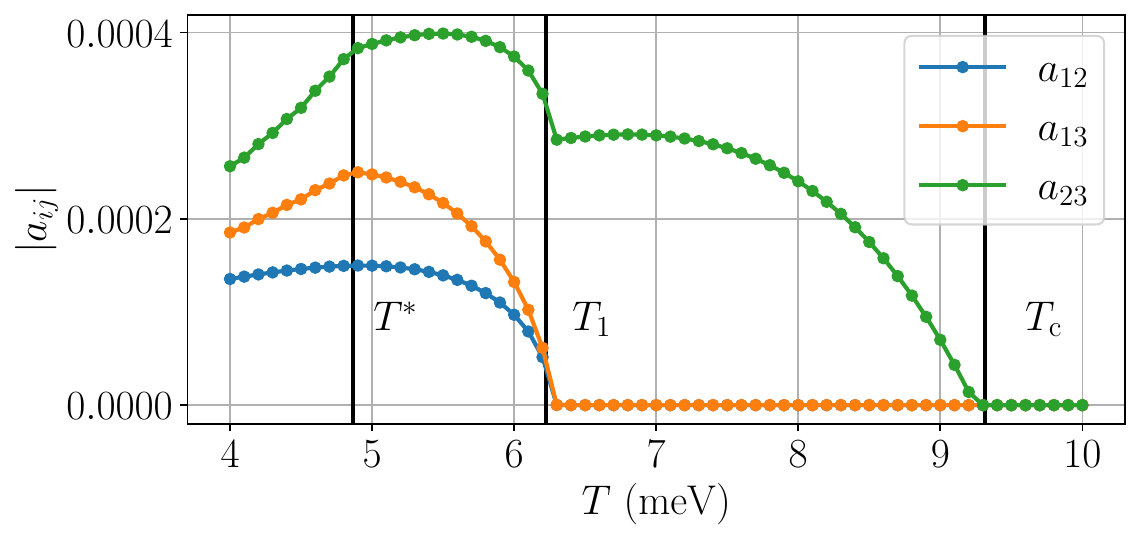}
    \caption{The quantities $a_{ij}$ defined in Eq. \eqref{a_ij_freq} as a function of temperature $T$. The system parameters and quantities are the same as in Fig.\ \ref{fig:temperature_dependence}. The black lines at $T=T^*$, $T=T_1$ and $T=T_\mathrm{c}$ are included to help guide the eye.}
    \label{fig:a_ij}
\end{figure}

To understand the phase transition at $T^*$, we repeat that there are two criteria for TRSB to occur: $a_{ij}$ must have the appropriate sign combinations, and $a_{ij}$ must be of similar magnitudes. The first criterion is satisfied since the $B^{-1}_{ij}$ decide the sign of $a_{ij}$. $B^{-1}_{13}$ and $B^{-1}_{23}$ are negative while $B_{12}^{-1}$ is positive, and thus $a_{ij}$ has a sign combination that allow for TRSB. We explain the second criterion in more detail in Appendix \ref{app:num}, but from the fact that $\theta_{13}$ and $\theta_{23}$ are zero above $T^*$ in Fig.\ \ref{fig:temperature_dependence}, we can infer the relative magnitudes of $a_{ij}$. Above $T^*$, $a_{23}$ and $a_{13}$ are larger than $a_{12}$ since it is energetically favorable to have $\theta_{13}=\theta_{23}=0$, leaving the $a_{12}\cos\theta_{12}$-term frustrated. As we approach $T^*$ from above, both $\Delta_1(\omega_n)$ and $\Delta_2(\omega_n)$ increase, in turn boosting $a_{12}$, as can be seen from Eq.\ \eqref{a_ij_freq}. Importantly, $a_{12}$ then grows relative to $a_{13}$ and $a_{23}$ since $\Delta_3(\omega_n)$ does not change much below $T_1$. So at $T^*$, $a_{12}$ becomes large enough for the phase frustration to manifest in TRSB, as all three phase differences start deviating from zero. This can be observed in detail in Fig.\ \ref{fig:a_ij}, where $|a_{ij}|$ is plotted as a function of temperature for the same system as in Fig.\ \ref{fig:temperature_dependence}. Moreover, in Appendix \ref{app:num}, we derive that for a system with equal $a_{13} = a_{23}=a$, TRSB occurs at $a_{12}>a/2$. From Fig. \ref{fig:a_ij} we find that a similar relation holds also for $a_{13} \neq a_{23}$, namely that TRSB occurs at $a_{12} \approx (a_{13} + a_{23})/4$.

All three phase differences onsets similarly at $T^*$ with critical exponents $\beta\approx 0.5$, with different prefactors. This is a manifestation of the nature of the phase transition, namely that a $Z_2$ symmetry is spontaneously broken along with TRS, resulting in an Ising transition. Any of the three $\theta_{ij}$ can be taken to be the associated order parameter. The Ising critical exponents are well-known in mean-field theory. Relevant to our system is $\beta=1/2$, thus corroborating the critical exponents we have found. 

\begin{figure}
    \centering
    \includegraphics[width=8.6cm,height=8.6cm,keepaspectratio]{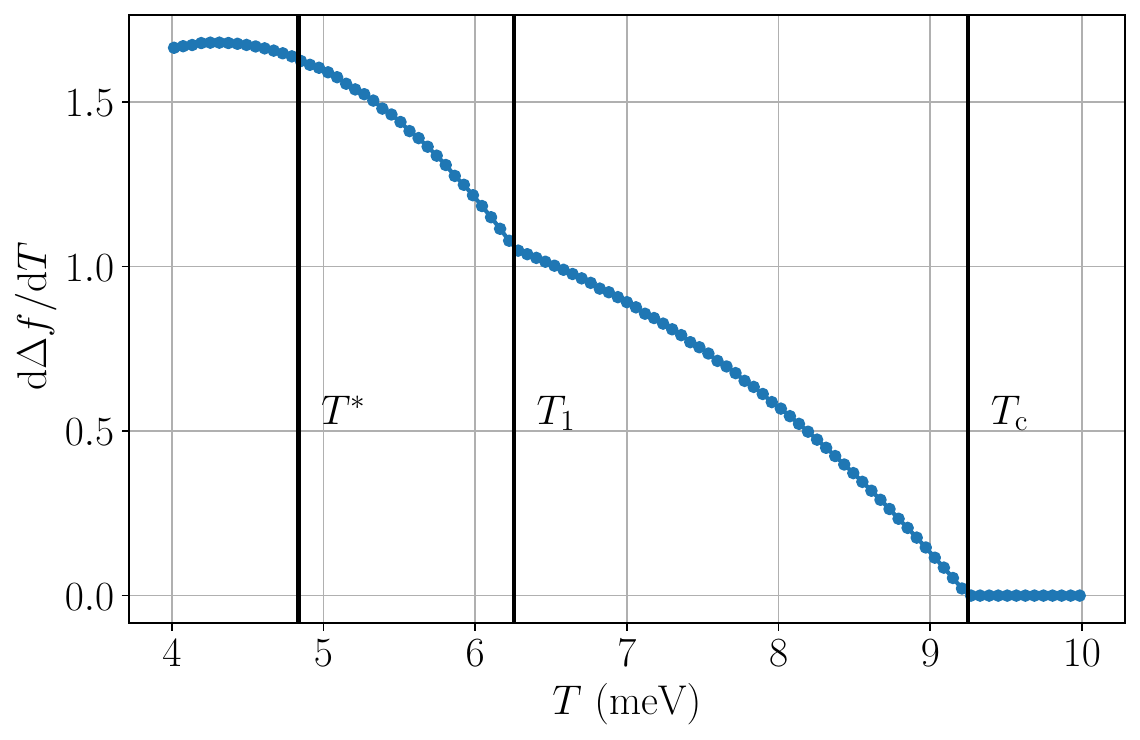}
    \caption{The derivative of the free energy density difference between the superconducting and normal state $\Delta f = f_{\mathrm{SC}}- f_\mathrm{N}$ with respect to temperature. The parameters and quantities are the same as in Fig.\ \ref{fig:temperature_dependence}. The black lines at $T=T^*$, $T=T_1$ and $T=T_\mathrm{c}$ are included to help guide the eye.}
    \label{fig:first_derivative}
\end{figure}

In superconductors with an additional phase transition below $T_\mathrm{c}$, one of its signatures is a jump in the heat capacity, which is most often associated with spontaneous TRSB \cite{Roising2022}. For multiband systems, the jump results from a kink in the derivative of the gaps at $T^*$ \cite{Wilson2013}. $\Delta_3(0)$ in Fig.\ \ref{fig:temperature_dependence} (b) clearly features such a kink, and with the mean-field Ising transition at $T^*$, we expect a logarithmic anomaly in the specific heat.
However, for the system considered here, the amplitude of the anomaly is minuscule due to the different magnitudes in the diagonal and off-diagonal elements in $a_{ij}$. $a_{11}$ and $a_{22}$ are five orders of magnitude larger than the remaining matrix elements, so although the off-diagonal elements in $a_{ij}$ determine $\theta_{ij}$ and thus play consequential roles in the emergent physics, their contribution to the free energy is insignificant. The dominant contributions from the gaps to the free energy are from $a_{11}$, $a_{22}$, and the tracelog contributions.
None of these terms contain $\theta_{ij}$, so when calculating the specific heat $c = T \mathrm{d}^2f/\mathrm{d}^2T$, the onset of $\theta_{ij}$ at $T^*$ does not contribute to any of the larger terms.

The absence of a kink at $T=T^*$ can be observed in Fig.\ \ref{fig:first_derivative}, where we have plotted $\mathrm{d} \Delta f/\mathrm{d}T$ as a function of $T$ with the same system parameters as in Fig.\ \ref{fig:temperature_dependence}. We calculate the difference between the superconducting and normal state $\Delta f = f_{\mathrm{SC}} - f_\mathrm{N}$ from Eq.\ \eqref{free_energy_with_only_freq}. Kinks in $\mathrm{d} \Delta f/\mathrm{d}T$ correspond to jumps in the heat capacity. In Fig.\ \ref{fig:first_derivative}, we observe that $\mathrm{d} \Delta f/\mathrm{d}T$, and thus also $c$, is mainly influenced by the temperature dependence of $|\Delta_1(\omega_n)|$ and $|\Delta_2(\omega_n)|$, resulting in two jumps in $c$ at $T_1$ and $T_\mathrm{c}$, associated with the onsets of $|\Delta_1(\omega_n)|$ and $|\Delta_2(\omega_n)|$, respectively. To obtain a noticeable kink at $T^*$, the off-diagonal elements in $a_{ij}$, and thus also the interband interactions, need to play a more prominent role. We consider such systems in the next paragraphs.

TRSB states are well understood in weak coupling. Hence, some comparisons to our results are in order. The conventional starting point for obtaining such states in BCS theory is systems with attractive intraband and repulsive interband interactions \cite{Stanev2010, Tanaka2010}. It then follows straightforwardly from the expression for $a_{ij}^{\mathrm{BCS}}$ that they are all positive, see Appendix \ref{app:parameters}, setting the stage for TRSB. However, from the interaction in Eq.\ \eqref{V_def}, it is obvious that the interaction we have employed is exclusively attractive, so such a sign change in the interband interactions must come from some other mechanism than electron-phonon coupling. 

Two potential candidates in this regard are the Coulomb interactions between the electrons or interactions mediated by spin fluctuations, where the latter are believed to be relevant in the superconducting iron pnictides \cite{Mazin2008, Hirschfeld2011}. These two can be incorporated into Eliashberg theory \cite{Ummarino2009}.
Their relative magnitudes, and thus also the overall sign in each interaction channel, are a matter of microscopic details. 
However, in the iron pnictides, the electron-phonon coupling is typically small in the interband channels \cite{Hirschfeld2011}, such that the other two interactions dominate the interband channels.
With the interband interactions playing a more prominent role, we also expect the anomaly in the specific heat at $T=T^*$ to be more notable, as discussed previously.
Moreover, when only including phonon-mediated interactions, the restrictions posed on the electron-phonon couplings leave some room for variations in $g_{ij}$ (See Appendix \ref{app:parameters} for details), but with additional scattering mechanisms, the parameter space resulting in TRSB will grow. 
Therefore, studying unconventional superconductors where mechanisms other than electron-phonon coupling may induce superconductivity constitutes another path that might lead to TRSB in strong coupling. We leave this for future investigations.

We lastly note that the different interactions considered in the previous paragraph are all easily integrated into Eliashberg theory using modern first-principles calculations \cite{Das2023}.
Such a strong-coupling treatment, grounded in first-principles calculations, would be interesting to apply to doped \ch{BaFe_2As_2} as it might shed some more light on the mechanisms causing the TRSB observed experimentally \cite{Grinenko2020} or to search for other candidate materials.

\section{Conclusions}\label{sec:conclusion}
In this paper, we have studied multiband superconducting states that spontaneously break time-reversal symmetry. These states have been intensely researched in the previous decade following their initial theoretical discovery \cite{Ng2009, Stanev2010, Tanaka2010}, but also in the last few years with experimental realizations \cite{Grinenko2020, Grinenko2021} of TRSB states. So far, most theoretical descriptions have employed either weak-coupling BCS theory or effective Ginzburg-Landau theories derived thereof. In this work, on the other hand, we have presented a strong-coupling analysis of such systems. 

Our most important finding is that the TRSB states may indeed emerge in strong-coupling models. They are more elusive than in weak coupling, as the time-reversal symmetric states also satisfy the Eliashberg equations that fixed-point schemes tend to favor. To compensate for this, we have derived the free energy of the system and used it to guide the numerical solver towards TRSB states when they were more energetically favorable than their TRS counterparts.
In particular, we have considered a system where two large gaps are weakly coupled to each other and a significantly smaller third gap. Although the third gap is small, the phase frustration in the system still drives the spontaneous breaking of TRS either at a critical electron-phonon coupling or at $T=T^*$.
We have discussed how one expects the emergent physics to be similar to that of TRSB three-band systems with equal gap amplitudes, namely spontaneous currents and magnetization arising near defects \cite{Lin2016, Garaud2017}.

The strong-coupling approach we have used is a generalization of the recent work of Ref.\ \cite{Protter2021} to multiband systems. Here, the Eliashberg equations emerge as stationary point conditions on the action in the imaginary-time functional-integral formalism. Moreover, this approach admitted a calculation of the free energy of the system in strong coupling,which is a quantity that plays a pivotal role in several aspects.  
As mentioned, we have used it to evaluate whether the TRSB and TRS states had the lowest free energy, as they both satisfy the stationary point conditions. Thus, the free energy provides a necessary selection criterion for choosing between the states. 
Furthermore, from the free energy, we have found that the energy associated with the phase frustration was small. Hence we neither expect nor observe an anomaly in the specific heat at $T=T^*$. 
We have also discussed other systems and scattering mechanisms that we expect to amplify the amplitude of the logarithmic anomaly in the specific heat. In particular, we have suggested that unconventional multiband superconductors may constitute a class of systems where TRSB states may emerge.

Finally, we have compared our results for TRSB in strong coupling with BCS theory. Although the gap equation is more numerically demanding in the former, the resulting TRSB states are still similar; the spatial symmetries of the gaps are the same ($s$-wave), and there is no frequency dependence in their phases, only in their magnitude. However, there are several benefits of employing strong coupling. The renormalization effects of the electrons are accounted for, and materials with large electron-phonon couplings can be investigated. 
Moreover, Eliashberg theory has the advantage of being well incorporated with first-principles calculations. So, the framework for studying TRSB in strong coupling presented here may apply these directly, thus facilitating a more effective search for candidate materials.

\section*{Acknowledgments}
We thank Christian Svingen Johnsen and Sondre Duna Lundemo for useful discussions. This work was supported by the Research Council of Norway (RCN) through its Centres of Excellence funding scheme, Project No. 262633, ``QuSpin'', as well as RCN Project No. 323766.

\appendix

\section{Numerical details on solving the Eliashberg equations}\label{app:num}
In this appendix, we will briefly cover how we solve the three-band Eliashberg equations in Eqs.\ \eqref{finished_eliash_eq_for_Z} and \eqref{finished_eliash_eq_for_Delta}. They consist of a set of coupled equations with $3N_\mathrm{M}$ complex variables (or, equivalently, $6N_\mathrm{M}$ real variables) for $\Delta_i(\omega_n)$, and $3N_\mathrm{M}$ real variables for $Z_i(\omega_n)$, where $N_\mathrm{M}$ is the number of Matsubara frequencies we include. In producing our results, we employed a Matsubara cutoff $M = 1100$ meV large enough to capture the physics at play. We checked that the results did not change when increasing $M$ further. $N_\mathrm{M}$ is given by the number of Matsubara frequencies that satisfy $|\omega_n|\leq M$.

With the global phase ansatz, the number of variables in $\Delta_i(\omega_n)$ is reduced to $3N_\mathrm{M} + 3$. Moreover, since only the phase differences $\theta_{ij}$ are present in Eq.\ \eqref{finished_eliash_eq_for_Delta}, we can reduce the number of variables to $3N_\mathrm{M} + 2$ since $\theta_{23} = \theta_{13} - \theta_{12}$. We will now demonstrate how the remaining phase differences $\theta_{12}$ and $\theta_{13}$ are determined. To accomplish this, as alluded to in the main text, we use the insight offered by the free energy. Specifically, the only term in the free energy dependent on $\theta_{12}$ and $\theta_{13}$ is given by
\begin{equation}
    a_{12}\cos\theta_{12} + a_{13} \cos\theta_{13} + a_{23} \cos(\theta_{13} - \theta_{12}),
    \label{phase_term_appendix}
\end{equation}
where the expression for $a_{ij}$ is in Eq.\ \eqref{a_ij_freq}. Because the solutions to the Eliashberg equations must satisfy stationary point conditions in the free energy, only values for $\theta_{12}$ and $\theta_{13}$ that extremize the free energy are admissible solutions. This is equivalent to extremizing Eq.\ \eqref{phase_term_appendix} above. 

There are, at maximum, six solutions to extremizing Eq.\ \eqref{phase_term_appendix} for $\theta_{ij} \; \in\;  [0, 2\pi)$. Four are TRS, namely \\
\noindent $(\theta_{12}, \theta_{13})= \{(0,0), (0,\pi), (\pi, 0), (\pi,\pi) \}$. The remaining two break TRS, and are time-reversal counterparts. Denoting one as $\theta^{\mathrm{TRSB}}_{ij}$, the other is equal to $2\pi-\theta^{\mathrm{TRSB}}_{ij}$. For $\theta^{\mathrm{TRSB}}_{ij}$ to exist, the quantity 
\begin{align}
    b &\equiv -a_{12}^4a_{13}^4 - a_{12}^4a_{23}^4 - a_{13}^4a_{23}^4 \nonumber\\
    &+ 2a_{12}^2a_{13}^2a_{23}^2(a_{12}^2+a_{13}^2+a_{23}^2) \label{b_def}
\end{align}
must be larger than zero. 
As a simple example, if $a_{12} = a_{13}\equiv a$ (which is physically realized if $g_{11}=g_{22}$ and $g_{13}=g_{23}$), $b>0$ corresponds to $a_{23}>a/2$. 

If $b>0$, $\theta^{\mathrm{TRSB}}_{ij}$ can be written in the form
\begin{align}
    \theta^{\mathrm{TRSB}}_{12} &= \mathrm{atan2}\bigg[ \frac{\sqrt{b}}{2a_{12}^2a_{13}a_{23}}, \frac{a_{13}^2a_{23}^2 - a_{12}^2a_{13}^2 + a_{12}^2a_{23}^2}{2a_{12}^2a_{13}a_{23}} \bigg] \\
    \theta^{\mathrm{TRSB}}_{13} &= \mathrm{atan2}\bigg[ \frac{\sqrt{b}}{2a_{12}a_{13}^2a_{23}}, \frac{a_{12}^2a_{23}^2 -a_{12}^2a_{13}^2 - a_{13}^2a_{23}^2}{2a_{12}a^2_{13}a_{23}} \bigg],
\end{align}
where $\mathrm{atan2}(x,y)$ is the two-argument inverse tangent. 
With the six solutions for $\theta_{12}$ and $\theta_{13}$, we can insert them into Eq.\ \eqref{phase_term_appendix} to determine which of them yields the lowest energy, thus offering an unambiguous method of determining the energetically favored phase configuration. 

Now, we can use the procedure above to solve the Eliashberg equations for the remaining $3 N_\mathrm{M}$ variables in both $Z_i(\omega_n)$ variables and $|\Delta_i(\omega_n)|$. Since we have in total $6 N_\mathrm{M}$ variables to solve for, it suffices to consider Eq.\ \eqref{finished_eliash_eq_for_Z} and the real part of Eq.\ \eqref{finished_eliash_eq_for_Delta}. Collecting the $6 N_\mathrm{M}$ variables in the vector $\mathbf{x}$, solving these equations is equivalent to fixed-point iteration in the form
\begin{equation}
    \mathbf{x}_{m+1} = g(\mathbf{x}_{m}),
    \label{fixed_point}
\end{equation}
where the function $g(\mathbf{x}_{m})$ constitutes the right-hand sides of Eqs. \eqref{finished_eliash_eq_for_Z} and \eqref{finished_eliash_eq_for_Delta}.
However, before $g(\mathbf{x}_n)$ is computed, based on the values for $Z_i(\omega_n)$ and $|\Delta_i(\omega_n)|$ in $\mathbf{x}_{m}$, we calculate $a_{ij}$ and subsequently determine the values of $\theta_{12}$ and $\theta_{13}$ in accordance with the procedure above. In this way, we ensure that we use the most energetically favorable $\theta_{ij}$ each time we evaluate the real part of Eq.\ \eqref{finished_eliash_eq_for_Delta} during the fixed-point scheme. Finally, we terminate the iteration in Eq.\ \eqref{fixed_point} once convergence (i.e.\ a prescribed tolerance level) has been reached. 

\section{Electron-phonon parameters}\label{app:parameters}
The purpose of this appendix is to elucidate some of the restrictions posed by the simple interaction we employ in Eq.\ \eqref{V_def} for the system to have TRSB and to motivate our choice of parameters in light of this. As stated in the main text, a necessary condition for TRSB is that either one or all of $a_{12}$, $a_{13}$, and $a_{23}$ must be positive. Since the sign of $a_{ij}$ is determined by $B_{ij}^{-1}$, it suffices to look at the general expression for off-diagonal elements for a $3\times 3$ matrix $A$
\begin{equation}
    A_{ij}^{-1} = \frac{1}{\det A}(A_{il}A_{jl} - A_{ij}A_{ll}),
    \label{A_inv_explicit}
\end{equation}
where $l\neq i\neq j$. Since $B_{ij}=|g_{ij}|^2$, all its elements are nonnegative, causing severe restrictions on the possible choices of $B_{ij}^{-1}$ that can cause phase frustration, as we will now see.

As discussed in Sec.\ \ref{sec:results}, in BCS theory, the conventional choice is to have attractive intraband and repulsive interband interactions, which straightforwardly yields all positive $[V^{\mathrm{BCS}}]_{ij}^{-1}$ from Eq.\ \eqref{A_inv_explicit}, supposing that the determinant of $V^{\mathrm{BCS}}$ is positive. In the model we employ, there is no way of having negative coupling constants in $B_{ij}$ unless one includes additional scattering mechanisms such as Coulomb repulsion and interactions mediated by spin fluctuations, as discussed in the main text.
However, positive values in $B_{ij}$ may still result in emergent TRSB, as demonstrated in e.g., Fig.\ \ref{fig:sols_with_freq_dep} (d). Since $g_{11}, g_{22}$ are much larger than the other $g_{ij}$, the same holds for $B_{11}$ and $B_{22}$. It then follows immediately from Eq.\ \eqref{A_inv_explicit} that $B_{13}^{-1}$ and $B_{23}^{-1}$ are negative. To obtain a positive $B_{12}^{-1} = (\det B)^{-1}(B_{13}B_{23} - B_{12}B_{33})$ while keeping $\det B$ positive, one possible choice is to have a minuscule value for $B_{12}$, which is the choice made in the main text. If it is zero, we checked that the results did not change qualitatively.

Interband electron-phonon couplings are small in iron pnictides \cite{Hirschfeld2011}, motivating our choice of using small values for $g_{ij}, \; i\neq j$. Moreover, with our main interest lying in the strong-coupling regime, we choose large values for $g_{11}$ and $g_{22}$. However, while there is a physical mechanism for the third intraband coupling $g_{33}$ to be small (see main text), the motivation for this is not immediately obvious. 
It is a choice made to satisfy $b>0$, where $b$ is given in Eq.\ \eqref{b_def}. If $b<0$, TRS cannot be spontaneously broken. To understand why $b>0$ requires a small $g_{33}$, one must consult the expression for $a_{ij}$ in Eq.\ \eqref{a_ij_freq}: Since $B_{12}^{-1}$ is orders of magnitudes smaller than $B_{13}^{-1}$ and $B_{23}^{-1}$, $|\Delta_1(\omega_n)|$ and $|\Delta_2(\omega_n)|$ must be approximately the same number of orders of magnitudes larger than $|\Delta_3(\omega_n)|$ for the amplitudes of $a_{12}$, $a_{13}$, and $a_{23}$ to satisfy $b>0$. Since the gaps grow with their respective intraband couplings, this is accomplished by having $g_{33}\ll g_{11}, g_{22}$.

\bibliography{main.bib}

\end{document}